\documentclass{article}
\usepackage{graphicx}
\usepackage{authblk}
\usepackage[a4paper]{geometry}
\usepackage{hyperref}
\usepackage{amsmath}
\usepackage{bm}
\usepackage{natbib}
\usepackage{amssymb}
\usepackage{subcaption}

\def\T{{ \mathrm{\scriptscriptstyle T} }}

\newcommand{\ind}{\mathbb{I}}

\newcommand{\btheta}{\bm{\theta}}
\newcommand{\bbeta}{\bm{\beta}}
\newcommand{\bnu}{\bm{\nu}}
\newcommand{\bmu}{\bm{\mu}}

\newcommand{\bzeta}{\bm{\zeta}}

\newcommand{\bvarrho}{\bm{\varrho}}

\newcommand{\Var}{\ensuremath \text{var}}
\newcommand{\R}{\mathbb{R}}

\newcommand{\gid}{\ensuremath \xrightarrow{\hspace*{0.08cm} d \hspace*{0.08cm}}}
\newcommand{\gip}{\ensuremath \xrightarrow{P}}

\newcommand{\Cov}{\ensuremath\text{cov}}

\newcommand{\E}{\ensuremath\mathnormal{E}}
\newcommand{\eqd}{\overset{d}{=}}

\DeclareMathOperator*{\argmax}{arg\,max}

\numberwithin{equation}{section}

\newtheorem{theorem}{Theorem}
\newtheorem{corollary}{Corollary}
\newtheorem{lemma}{Lemma}
\newtheorem{assumption}{Assumption}
\newtheorem{remark}{Remark}

\newcounter{example}[section]
\newenvironment{example}[1][]{\refstepcounter{example}\par\medskip
   \noindent \textit{Example~\theexample ~(#1).} \rmfamily}{\medskip}

\title{
    Sample Splitting and Assessing Goodness-of-fit of Time Series
}
\author[1]{Richard A. Davis}
\author[1]{Leon Fernandes}
\affil[1]{\small\textit{Department of Statistics, Columbia University}}

\begin{document}

\maketitle

\begin{abstract}
    A fundamental and often final step in time series modeling is to assess the quality of fit of a proposed model to the data.
    Since the underlying distribution of the innovations that generate a model is often not prescribed, goodness-of-fit tests typically take the form of testing the fitted residuals for serial independence.
    However, these fitted residuals are inherently dependent since they are based on the same parameter estimates and thus standard tests of serial independence, such as those based on the autocorrelation function (ACF) or distance correlation function (ADCF) of the fitted residuals need to be adjusted.
    The sample splitting procedure in Pfister et al.~(2018) is one such fix for the case of models for independent data, but fails to work in the dependent setting.
    In this paper sample splitting is leveraged in the time series setting to perform tests of serial dependence of fitted residuals using the ACF and ADCF.
    Here the first $f_n$ of the data points are used to estimate the parameters of the model and then using these parameter estimates, the last $l_n$ of the data points are used to compute the estimated residuals.
    Tests for serial independence are then based on these $l_n$ residuals.
    As long as the overlap between the $f_n$ and $l_n$ data splits is asymptotically 1/2, the ACF and ADCF tests of serial independence tests often have the same limit distributions as though the underlying residuals are indeed iid.
    In particular if the first half of the data is used to estimate the parameters and the estimated residuals are computed for the entire data set based on these parameter estimates, then the ACF and ADCF can have the same limit distributions as though the residuals were iid.
    This procedure ameliorates the need for adjustment in the construction of confidence bounds for both the ACF and ADCF in goodness-of-fit testing. 
\end{abstract}


\section{Introduction}

One of the most important steps in the process of fitting models is to verify a model prior to making final inferences about the dataset.
Fitting a model to data necessitates goodness-of-fit tests to assess the aptness and quality of the model fit.
In a time series setting, many models are formulated as functions of an underlying iid noise process.
Looking at the residuals, which are estimates of the noise sequence, and performing autocorrelation and serial independence tests to ascertain whiteness of the residual process is the standard approach for goodness-of-fit.
However these tests cannot be applied as though the residuals are uncorrelated or independent.
For autoregressive models \citet{BoxPierce1970} exhibits how the asymptotic behavior of the autocorrelation function (ACF) depends on the lag.
Again for serial dependence testing using distance correlation function (ADCF), Theorem 3.1 of \citet{DavisWan2020} and Theorem 4.1 of \citet{Davisetal2018} include a correction term $\xi_h$ to the limit distribution which would be absent if the residuals were actually independent.
This problem of not knowing the null distributions for the residuals limits the usefulness of these tests, and require adjustments in order to be applied correctly. 

To alleviate this problem we propose a simple sample splitting procedure which offers the opportunity of residuals which behave as though they are iid when applying these goodness-of-fit tests.
These residuals are obtained by splitting the data as follows: (i) the analysis split $\{X_1, \ldots, X_{f_n}\}$ is used to estimate $\bbeta$, (ii) the assessment split $\{X_{n - l_n + 1}, \ldots, X_n\}$ is used to calculate the residuals based on the estimate $\hat{\bbeta}$.
We let the sample splitting sequences be arbitrary non-decreasing sequences $\{(f_n, l_n)\}_{n \geq 1}$ which go to infinity with $n$ such that $f_n/n \rightarrow k_f$ and $l_n/n \rightarrow k_l$ for constants $0 < k_f, k_l \leq 1$.

Our results are stated for a general class of models and corresponding estimation procedures which are then verified for ARMA and GARCH models.
In either case we show that the sample split $f_n = n / 2$ and $l_n = n$ yields residuals with the same asymptotic properties for the goodness-of-fit tests as that of iid noise.
For example the residual ACF will be asymptotically iid standard normal at various lags.
This immediately implies that a simple chi-squared test can be performed on the ACF process of the residuals for assessing goodness-of-fit.
Our method bypasses dependence accounting adjustments such as in the Box-Pierce and Ljung-Box tests \citep{BoxPierce1970,LjungBox1978}.

The half-sample device of \citet{Durbin1973} is similar to our method in the context of Kolmogorov-Smirnov goodness-of-fit type tests based on iid data.
Here half the data is used for estimation and then the entire dataset is used for evaluation; it is shown that the asymptotic properties of the sample distribution so obtained is the same as when the true parameter is known.
For further details see for example \citet{Durbin1976, Stephens1978, Stephens1986} and the references therein.

A crucial step in our method is to allow for overlap between the two splits.
It may initially seem convenient to take disjoint subsets with $f_n + l_n = n$ to get independence.
In fact for the dHSIC independence test in \citet{Pfisteretal2018}, a 50-50 sample split for goodness-of-fit purposes is employed in an independent data setting.
However this method fails in a time series setting because of inherent dependence of each observation to its past.
In fact we will show that using disjoint splits exacerbates the problem -- the asymptotic distribution of the iid noise process for the ACF will always be stochastically dominated by the corresponding asymptotic distribution for the residuals.

Section~\ref{sec-acf} starts with an empirical demonstration of the results for the ACF applied to sample splitting based residuals.
The first main result in Section~\ref{sec-acf-gen} develops the theory applied to general time series models. 
For applications of the general results to specific models, we study sample splitting for ARMA and GARCH models in Sections~\ref{sec-acf-arma} and \ref{sec-acf-garch} respectively.
Section~\ref{sec-adcf}  contains the second main result which discusses sample splitting with the ADCF as a measure of serial dependence rather than the ACF.
A simulation study is performed to verify the results.

\section{Asymptotics for ACF of Residuals}
\label{sec-acf}

\subsection{Autoregressive model}
\label{sec-acf-ar1}

We illustrate the basic idea in the case of an autoregressive model $X_j = \beta X_{j - 1} + Z_j$, where $\{Z_j\}$ is an iid sequence with $\E[Z_0] = 0$ and $\Var[Z_0] = \sigma^2 < \infty$.
Let the least squares estimate $\hat{\beta}_{f_n}$ be based only the first $f_n$ observations.
The sample splitting residuals are based on the last $l_n$ observations are $\hat{Z}_{n - l_n + 1}, \ldots, \hat{Z}_n$, where the residual at time $j$ is $\hat{Z}_j = X_j - \hat{\beta}_{f_n} X_{j - 1}$.
The ACF of the residuals at some lag $h > 0$ is then
\begin{align*}
    \hat{\rho}_{l_n}^{\hat{Z}}(h) = \frac{\sum_{j = n - l_n + 1}^{n - h} \hat{Z}_j \hat{Z}_{j + h}}{\sum_{j = n - l_n + 1}^{n} \hat{Z}_j^2}.
\end{align*}
For the case $f_n = l_n = n$, Example 9.4.1 of \citet{BrockwellDavis1991} shows that the ACF is asymptotically Gaussian with mean zero and variance $1 - \beta^{2(h - 1)}(1 - \beta^2)$.

Our first result is used to obtain the sample splitting residual ACF as asymptotically Gaussian with mean zero and variance $1 + (k_{ra} - 2k_{ov})\beta^{2(h - 1)}(1 - \beta^2)$; see Remark~\ref{rmk-ar1} for details.
Here $k_{ra} = \lim_{n \rightarrow \infty} {l_n}/{f_n}$ is the limiting ratio and $k_{ov} = \lim_{n \rightarrow \infty} {\max\{0, f_n + l_n - n\}}/{f_n}$ is the limiting overlap coefficient of the sample splitting sequences.
The sign of $(k_{ra} - 2k_{ov})$ determines if the asymptotic residual ACF is stochastically smaller or larger than the ACF values for the true noise.
To want them equal we need to ensure $k_{ra} = 2k_{ov}$ and the most obvious choice for this is with $f_n = n/2$ and $l_n = n$.

\begin{figure}[t]
    \centering
    \includegraphics[width=0.8\textwidth]{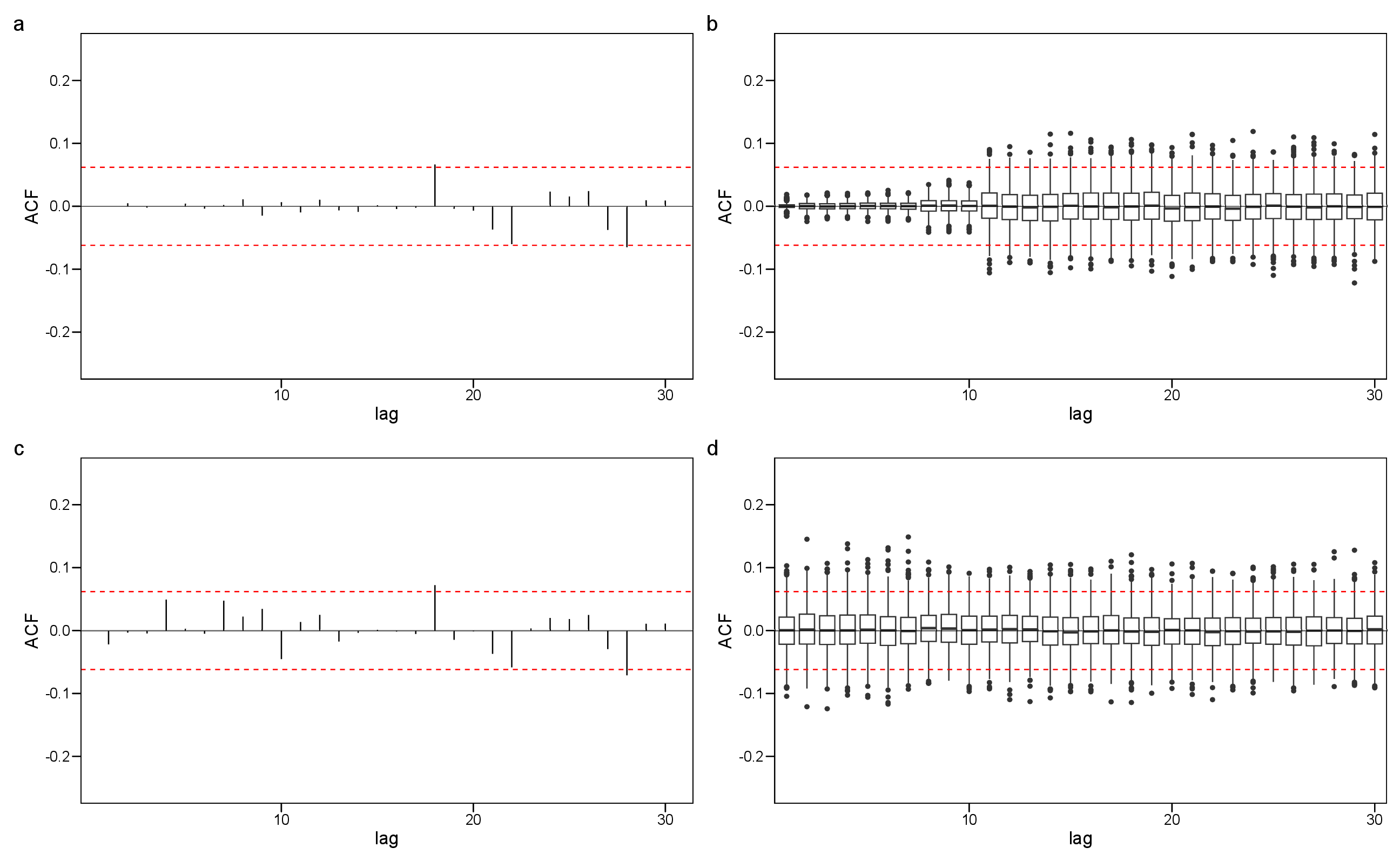}
    \caption{
        Residual ACF plots for AR(10) model at lags 1 to 30.
        Panels on the left side are for a single simulation, while panels on the right side are the boxplots of the ACF obtained based on 1000 simulations.
        Each simulation uses $n = 1000$ with the top row corresponding to $f_n = n$ and bottom row corresponding to $f_n = n / 2$, with $l_n = n$ in either case.
        Dashed red line is 95\% quantile of standard Gaussian, that is $\pm 1.96/\sqrt{1000}$.
    }
    \label{fig-ar10-normal-demo}
\end{figure}

To demonstrate this result, as well as its generality, we consider the residual ACF of an autoregressive model of order 10 with AR coefficients
\begin{align*}
    \bbeta = (-0.140, 0.038, 0.304, 0.078, 0.069, 0.013, 0.019, 0.039, 0.148, -0.062).
\end{align*}
The innovations used are iid standard Gaussian to simulate data of length $n = 1000$ from the above model.
Using least squares we fit an AR(10) model and obtain the sample splitting based residuals corresponding to two cases $f_n = n/2$ and $f_n = n$, with $l_n = n$ for either case.
The ACFs for these residuals are shown in Figure~\ref{fig-ar10-normal-demo}.
Notice the really small values among the first 10 lags in Figure~\ref{fig-ar10-normal-demo} (a) and Figure~\ref{fig-ar10-normal-demo} (b).
This demonstrates the deviation of the residual ACF from a standard normal under $f_n = n$.
The ACF values with $f_n = n/2$ in Figure~\ref{fig-ar10-normal-demo} (c) and (d) are significantly higher at the initial lags and can be correctly be compared with the standard Gaussian quantile at each lag.

\subsection{General framework}
\label{sec-acf-gen}

We will study the asymptotics under a very general class of models.
Let $X_1, \ldots, X_n$ be observations from the parametric model given by
\begin{align}
    \label{eq-model}
    X_j = g(Z_{-\infty:j}; \bbeta_0),
\end{align}
where $\bbeta_0 \in \mathbf{\Theta}$ for some compact set $\mathbf{\Theta} \subset \R^p$, and $\{Z_j\}_{j \in \mathbb{Z}}$ is iid with mean zero and finite variance $\sigma^2$.
Assuming the model is invertible we have
\begin{align}
    \label{eq-invert}
    Z_j = h(X_{-\infty:j}; \bbeta_0).
\end{align}
For a model parameter $\bbeta$, we write $Z_j(\bbeta) := h(X_{-\infty:j}; \bbeta)$.
If $\bbeta^\dagger$ is an estimate of $\bbeta$, then the unobserved residual at time $j$ is
\begin{align*}
    \tilde Z_j := Z_j(\bbeta^\dagger) = h(X_{-\infty:j}; \bbeta^\dagger)
\end{align*}
Since $X_j$ is not observed for $j\leq 0$, we obtain the residuals as
\begin{align*}
    \hat Z_j(\bbeta^\dagger) = h(Y_{-\infty:j}; \bbeta^\dagger)
\end{align*}
where $Y_j = X_j,\, j \geq 1$ and $Y_j = 0 ,\, j \leq 0$.
Before stating the main result, we discuss the relevant assumptions.
\begin{assumption}
    \label{cond-m-est}
    Let $\mathcal{F}_j$ be the $\sigma$-algebra generated by $\{X_k, k \leq j\}$.
    We assume that the parameter estimate $\hat{\bbeta}_n$ is of the form
    \begin{align}
        \label{eq-m-est}
        \sqrt{n}(\hat{\bbeta}_n - \bbeta_0) = \frac{1}{\sqrt{n}} \sum_{j=1}^{n} \mathbf{m}(Z_{-\infty:j};\bbeta_0) + o_P(1),
    \end{align}
    where $\mathbf{m}$ is a vector-valued function of $X_{-\infty:j}$ such that 
    \begin{align}
        \label{eq-m-moments}
        \E [\mathbf{m}(Z_{-\infty:j};\bbeta_0)|\mathcal{F}_{j-1}] = \mathbf0, \quad \E|\mathbf{m}(Z_{-\infty:0};\bbeta_0)|^2 < \infty.
    \end{align}
    By the martingale central limit theorem, \eqref{eq-m-est} and \eqref{eq-m-moments} imply that
    \begin{align}
        \label{eq-m-asymp-q}
        \sqrt{n}(\hat{\bbeta}_n - \bbeta_0) \gid \mathbf{Q},
    \end{align}
    where $\mathbf{Q}$ is a random Gaussian vector with mean zero and variance 
    \begin{align*}
        \E[\mathbf{m}(Z_{-\infty:0}; \bbeta_0) \mathbf{m}(Z_{-\infty:0}; \bbeta_0)^\T].
    \end{align*}
\end{assumption}
\begin{assumption}
    \label{cond-l}
    Assume that the function $h$ in the invertible representation \eqref{eq-invert} is continuously differentiable, and writing
        \begin{align*}
            \mathbf{L}_j(\bbeta) := \frac{\partial}{\partial\bbeta} h(X_{-\infty:j};\bbeta),
        \end{align*}
        we assume for some $\delta_0>0$ that
        \begin{align}
            \label{eq-L-moments}
            \E \sup_{\|\bbeta-\bbeta_0\| \leq \delta_0} \| \mathbf{L}_0 (\bbeta) \|^2 < \infty.
        \end{align}
\end{assumption}
\begin{assumption}
    \label{cond-tilde-z}
    For any sequence of estimators $\bbeta_n^\dagger$ with $|\bbeta_n^\dagger- \bbeta_0| = O_P(n^{-1/2})$,
        \begin{align*}
            \frac{1}{\sqrt{n}} \sum_{j = 1}^n |\hat{Z}_{j}(\bbeta_n^\dagger) - \tilde{Z}_{j}(\bbeta_n^\dagger)|^k = o_P(1),\, k = 1, 2.
        \end{align*}
\end{assumption}

Although stated more generally, Assumption~\ref{cond-tilde-z} will be applied to the sequence of sample splitting based estimators $\hat{\bbeta}_{f_n}$, the estimated parameter based on the first $f_n$ observations.
The sample splitting based residuals are $\hat{Z}_{n - l_n + 1}, \ldots, \hat{Z}_n$, where $\hat{Z}_j = \hat{Z_j}(\hat{\bbeta}_{f_n})$ at time $j$.
For a fixed lag $h > 0$ we show that $l_n^{-1/2} \hat{\rho}_{l_n}^{\hat{Z}}(h)$ is asymptotically normal with mean 0 and variance
\begin{align}
    \label{eq-fin-acf-var}
    \sigma^2_h = 1 + \sigma^{-4} (2k_{ov} \bmu_{1 h}^\T \bmu_{2 h} + k_{ra} \bmu_{1 h}^\T \Var(\mathbf{Q}) \bmu_{1 h}),
\end{align}
where $\bmu_{1 h} = \E[ Z_0 \mathbf{L}_{h}(\bbeta_0)]$ and $\bmu_{2 h} = \E[\mathbf{m}(Z_{-\infty:{h}}; \bbeta_0) Z_0 Z_{h}]$.
This expression elucidates how the model affects the residuals and alters the variance away from 1, the autocorrelation of the true noise $Z_t$.
The result extends to joint convergence across lags and we show in the following that $\sqrt{l_n} (\hat{\rho}^{\hat{Z}}_{l_n}(1), \ldots, \hat{\rho}^{\hat{Z}}_{l_n}(h))^\T$ is asymptotically Gaussian with mean zero and covariance matrix
\begin{align}
    \label{eq-fin-acf-var-mat}
    \Sigma_h :=
    I_h + \sigma^{-4} (k_{ov} M_1 M_2^\T + k_{ov} M_2 M_1^\T + k_{ra} M_1 \Var(\mathbf{Q}) M_1^\T),
\end{align}
where
\begin{align*}
   M_1 = \begin{pmatrix}
    \bmu_{1 1}^\T \\
    \vdots \\
    \bmu_{1 h}^\T
   \end{pmatrix}_{h \times p} \;\; \text{ and } \;\;
   M_2 = \begin{pmatrix}
    \bmu_{2 1}^\T \\
    \vdots \\
    \bmu_{2 h}^\T
   \end{pmatrix}_{h \times p}.
\end{align*}
\begin{theorem}
    \label{thm-fin-acf-asymp}
    Let $X_1, \ldots, X_n$ be observations from the model \eqref{eq-model} and $\hat{\bbeta}_n$ be an estimator of $\bbeta$ satisfying Assumptions~\ref{cond-m-est}--\ref{cond-tilde-z}.
    For a sample splitting sequence $\{(f_n, l_n)\}_{n \geq 1}$, write $\hat{Z}_j = \hat{Z}_j(\hat{\bbeta}_{f_n})\ (j = n - l_n + 1, \ldots, n)$.
    Then for all $h > 0$
    \begin{align*}
        \sqrt{l_n} \big( \hat{\rho}^{\hat{Z}}_{l_n}(1), \ldots, \hat{\rho}^{\hat{Z}}_{l_n}(h) \big)^\T \gid \mathcal{N}_h (\mathbf{0}, \Sigma_h).
    \end{align*}
\end{theorem}

The covariance matrix \eqref{eq-fin-acf-var-mat} explicitly states the effect of the model and the estimation procedure on the limiting distribution.
As seen in \eqref{eq-fin-acf-var-mat}, both these effects may be influenced by the choice of the sample splitting sequences.
Consider the case where the splits are disjoint.
It is evident that when $k_{ov} = 0$ we have $(\Sigma_h - I_h)$ is nonnegative definite.
Thus the residual ACFs are stochastically larger in absolute value than the true noise ACFs for each corresponding lag.

Under $M_2 = -  M_1 \Var(\mathbf{Q})$, the covariance matrix simplifies to
\begin{align}
    \label{eq-fin-acf-var-cond}
    \Sigma_h &=
    I_h + \sigma^{-4} (k_{ra} - 2 k_{ov}) M_1 \Var(\mathbf{Q}) M_1^\T.
\end{align}
The sign of $(k_{ra} - 2k_{ov})$ decides whether or not the residual ACFs are stochastically larger or smaller than true noise ACFs; at $k_{ra} = 2k_{ov}$ we get identical asymptotics for the residual ACF and true noise ACF, that is $\Sigma_h = I_h$.
\begin{corollary}
    \label{thm-fin-acf-asymp-cond}
    Assume the conditions of Theorem~\ref{thm-fin-acf-asymp} and let $f_n = n / 2$ and $l_n = n$.
    Then for all $h > 0$, whenever $M_2 = - M_1 \Var(\mathbf{Q})$ we have
    \begin{align*}
        \sqrt{n} \big( \hat{\rho}^{\hat{Z}}_{n}(1), \ldots, \hat{\rho}^{\hat{Z}}_{n}(h) \big)^\T \gid \mathcal{N}_h(\mathbf{0}, I_h).
    \end{align*}
\end{corollary}
This corollary can be used for testing model adequacies in two ways.
Firstly, using sample splitting sequences as above yields residuals with ACF being independent standard Gaussians at different lags.
These ACF values will therefore be within the 95\% confidence interval $\pm 1.96/\sqrt{n}$.
Secondly, a chi-squared goodness-of-fit test can be performed to check the fit of the model.
Explicitly, the model is rejected when $n \sum_{k = 1}^{h} \big( \hat{\rho}_{n}^{\hat{Z}}(k) \big)^2$ is larger than $\chi^{2}_{h}(0.95)$, where $\chi^{2}_{h}(0.95)$ is the 0.95 quantile of a chi squared distribution with $h$ degrees of freedom.
Next we show that the assumptions of Corollary~\ref{thm-fin-acf-asymp-cond} are satisfied by ARMA models and GARCH models.

\subsection{ARMA model}
\label{sec-acf-arma}

We show in this section that ARMA models satisfy Assumptions~\ref{cond-m-est}--\ref{cond-tilde-z} with pseudo maximum likelihood estimation.
Consider a causal and invertible ARMA($p, q$) process given by
\begin{align*}
    X_j = \sum_{k = 1}^p \phi_k X_{j - k} + Z_j + \sum_{\ell = 1}^q \theta_{\ell} Z_{j - \ell},
\end{align*}
where $\bbeta_0 = (\phi_1, \ldots, \phi_p, \theta_1, \ldots, \theta_q)$.
Here $\{Z_j\}$ is an iid process with mean zero and variance $\sigma^2 < \infty$.
The above recursion can be written as
\begin{align*}
    \phi(B) X_j = \theta(B) Z_j,
\end{align*}
where $\phi(z) = 1 - \sum_{k = 1}^p \phi_k z^k$ is the AR polynomial, $\theta(z) = 1 + \sum_{\ell = 1}^q \theta_{\ell} z^{\ell}$ is the MA polynomial and $B$ is the backshift operator such that 
\begin{align*}
    B X_j = X_{j - 1}.
\end{align*}
Due to invertibility, $\phi(z) / \theta(z)$ has a power series expansion for which we can write
\begin{align*}
    \frac{\phi(z)}{\theta(z)} = \sum_{j = 0}^{\infty} \pi_j(\bbeta_0) z^j.
\end{align*}
The unobserved residual at time $j$ for an estimate $\bbeta^\dagger$ is given by
\begin{align*}
    \tilde{Z}_{j} := Z_{j}(\bbeta^\dagger) = \sum_{k = 0}^{\infty} \pi_k (\bbeta^\dagger) X_{j - k}.
\end{align*}
Thus the fitted residuals are found by truncating this sum at $j - 1$,
\begin{align*}
    \hat{Z}_{j}(\bbeta^\dagger) = \sum_{k = 0}^{j - 1} \pi_k (\bbeta^\dagger) X_{j - k}.
\end{align*}
We will consider the estimate $\hat{\bbeta}_n$ to be the pseudo maximum likelihood estimator based on Gaussian likelihood
\begin{align*}
    L(\bbeta, \sigma^2) \propto \sigma^{-n} (\det \Sigma)^{-1/2} \exp \Big( \frac{-1}{2\sigma^2} \mathbf{X}_{n}^\T \Sigma^{-1} \mathbf{X}_{n} \Big),
\end{align*}
where $\mathbf{X}_{n} = (X_1, \ldots, X_{n})^\T$ and the covariance $\Sigma(\bbeta) := \Var(\mathbf{X}_{n}) / \sigma^2$ is independent of $\sigma^2$.
Thus the pseudo maximum likelihood estimates are the values that maximize $L(\bbeta, \sigma^2)$.
It is shown in \citet{BrockwellDavis1991} that $\hat{\bbeta}_n$ is consistent and asymptotically normal even for non-Gaussian $Z_j$.

We can now verify the hypotheses of Corollary~\ref{thm-fin-acf-asymp-cond} for sample splitting based residual ACF.
Assumptions~\ref{cond-m-est}--\ref{cond-tilde-z} are verified in Section~\ref{app-arma-assump}.
From equations (8.11.5) - (8.11.8) in \citet{BrockwellDavis1991} we have
\begin{align*}
    \sqrt{n} (\hat{\bbeta}_n - \bbeta_0) = \frac{-1}{\sqrt{n}} \sigma^{-2} \Var(\mathbf{Q}) \sum_{j = 1}^{n} \mathbf{L}_j(\bbeta_0) Z_j + o_P(1),
\end{align*}
so that
\begin{align*}
    \mathbf{m}(Z_{-\infty:j}; \bbeta_0) = - \Var(\mathbf{Q}) \mathbf{L}_j (\bbeta_0) Z_j / \sigma^2.
\end{align*}
Then, $\bmu_{2 j} = - \Var(\mathbf{Q}) \E[\mathbf{L}_j (\bbeta_0) Z_0 Z_j^2] / \sigma^2 = - \Var(\mathbf{Q}) \bmu_{1 j}$.
Therefore $M_2 = - M_1 \Var(\mathbf{Q})$ follows.

\begin{figure}[t]
    \centering
    \includegraphics[width=0.8\textwidth]{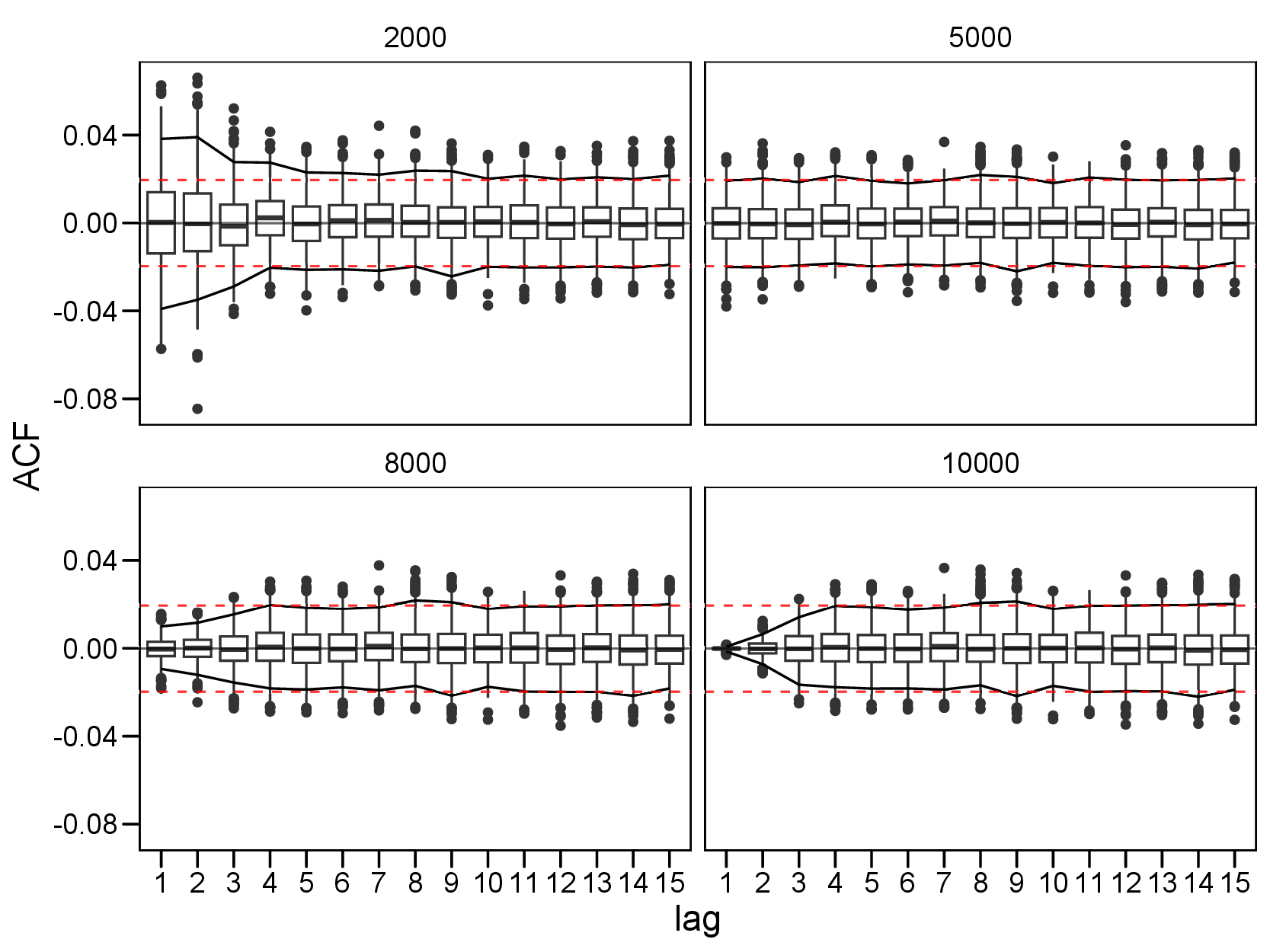}
    \caption{
        Boxplots of the residual ACFs at lags 1 to 15 for an ARMA$(2, 1)$ model with Laplace noise.
        Here $l_n$ is 10000 and each subplot corresponds to the value of $f_n$.
        The quantiles are obtained from 1000 simulations of length 10000.
        The solid lines are the 2.5\% and 97.5\% quantiles of the residual ACF and the dashed red lines denote corresponding quantiles of a standard Gaussian, that is $\pm 1.96/\sqrt{10000}$.
    }
    \label{fig-ar10-laplace-acf}
\end{figure}

We demonstrate this result for an ARMA$(2, 1)$ model.
In each simulation, a time series of length $n = 10000$ is generated from an ARMA$(2, 1)$ model with parameter
\begin{align*}
    \bbeta_0 = (\phi_1, \phi_2, \theta_1) = (0.8, 0.1, 0.3),
\end{align*}
where the innovations are Laplace with density function $f(z) = e^{-\sqrt{2} |z|} / \sqrt{2}$.
To illustrate the results of sample splitting, four different splits are considered wherein $l_n = 10000$ and $f_n$ is one of 2000, 5000, 8000, 10000.
For each simulated time series, we calculate the residual ACF for each sample split.
This is repeated 1000 times and we plot the 95\% confidence interval for the distribution of the ACF in Figure~\ref{fig-ar10-laplace-acf}, where each panel corresponds to the value of $f_n$.
The top row with $f_n \leq 5000$ corresponds to $k_{ra} \geq 2k_{ov}$ and we indeed see the asymptotic distribution of the ACF being larger than the corresponding standard Gaussian quantiles $\pm 1.96 / \sqrt{n}$.
Similarly the bottom row corresponds to $k_{ra} < 2k_{ov}$ and in this regime we see the asymptotic distribution of the ACF being stochastically smaller than the corresponding standard Gaussian quantiles.
As predicted under Corollary~\ref{thm-fin-acf-asymp-cond}, at $f_n = 5000$ the distribution of the ACF corresponds to standard Gaussian at each lag.

\begin{remark}
    \label{rmk-ar1}
    For the AR(1) model in the beginning of this section given by $X_t = \beta_0 X_{t-1} + Z_t$, we have $h(X_{-\infty:j}; \beta) = X_j - \beta X_{j - 1}$. Then $\mathbf{L}_{j}(\beta) = - X_{j - 1}$.
    Since the least squares estimator $\hat{\beta}_n$ satisfies
    \begin{align*}
        \hat{\beta}_n - \beta_0 =  \frac{1}{\Var(X_0)} \frac{1}{n} \sum_{j = 2}^{n} X_{j - 1} Z_j  + o_P(1),
    \end{align*}
    we have
    \begin{align*}
        \mathbf{m}(Z_{-\infty:j}; \beta_0) &= X_{j - 1} Z_j / \Var(X_0) = (1 - \beta_0^2) X_{j - 1} Z_j / \sigma^2, \\
        \Var(\mathbf{Q}) &= \E[\mathbf{m}(Z_{-\infty:j}; \beta_0)^2] =  (1 - \beta_0^2), \\
        \bmu_{1j}  &= - \E[Z_0 X_{j - 1}] = - \sigma^2 \beta_0^{j - 1},\\
        \bmu_{2j} &= (1 - \beta_0^2) \E[X_{j - 1} Z_0 Z_h^2] / \sigma^2 = - \Var(\mathbf{Q}) \bmu_{1j}.
    \end{align*}
    Thus the asymptotic covariance of sample splitting based residual ACFs at lags $j$ and $k$ is
    \begin{align*}
        \ind\{j = k\} + \sigma^{-4} (k_{ra} - 2k_{ov}) \bmu_{1j} \Var(\mathbf{Q}) \bmu_{1k} = \ind\{j = k\} + (k_{ra} - 2k_{ov}) \beta_0^{j + k - 2} (1 - \beta_0^2).
    \end{align*}
    In particular $1 + (k_{ra} - 2k_{ov})\beta_0^{2(j-1)}(1-\beta_0^2)$ is the asymptotic variance of the residual ACF.
\end{remark}

\subsection{ACF of squared residuals}
\label{sec-acf-sq-gen}

To check fit of GARCH models, the squared residual ACF is a popular choice.
In this section we briefly investigate the asymptotics of this statistic for sample splitting based residuals in the general framework of Section~\ref{sec-acf-gen}.
The squared residual ACF is given by
\begin{align*}
    \hat{\rho}_{l_n}^{\hat{Z}^2}(h) = \frac{\sum_{j = n - l_n + 1}^{n - h} \hat{Z}_j^2 \hat{Z}_{j + h}^2 - l_n^{-1} (\sum_{j = n - l_n + 1}^{n} \hat{Z}_j^2)^2}{\sum_{j = n - l_n + 1}^{n} \hat{Z}_j^4 - l_n^{-1} (\sum_{j = n - l_n + 1}^{n} \hat{Z}_j^2)^2}.
\end{align*}

The necessary modification of Assumption~\ref{cond-tilde-z} is stated below.
\begin{assumption}
    \label{cond-tilde-z-2}
    For any sequence of estimators $\bbeta_n^\dagger$ with $|\bbeta_n^\dagger- \bbeta_0| = O_P(n^{-1/2})$,
        \begin{align*}
            \frac{1}{\sqrt{n}} \sum_{j = 1}^n |\hat{Z}_{j}^2 (\bbeta_n^\dagger) - \tilde{Z}_{j}^2 (\bbeta_n^\dagger)|^k = o_P(1), \, k = 1, 2.
        \end{align*}
\end{assumption}

We obtain a very similar result in this case where $\sqrt{l_n} (\hat{\rho}^{\hat{Z}^2}_{l_n}(1), \ldots, \hat{\rho}^{\hat{Z}^2}_{l_n}(h))^\T$ is asymptotically Gaussian with mean zero and covariance matrix
\begin{align*}
    \tilde{\Sigma}_h :=
    I_h + (k_{ov} \tilde{M}_1 \tilde{M}_2^\T + k_{ov} \tilde{M}_2 \tilde{M}_1^\T + k_{ra} \tilde{M}_1 \Var(\mathbf{Q}) \tilde{M}_1^\T) / (\E[Z_0^4] - \sigma^4)^2,
\end{align*}
where $\tilde{M}_1$ and $\tilde{M}_2$ are $h \times p$ matrices with $k$th rows given respectively by
\begin{align*}
    \bmu_{1 k}^{(2)} &= 2 \E[Z_0^2 Z_k \mathbf{L}_k(\bbeta_0) - \sigma^2 Z_0 \mathbf{L}_0(\bbeta_0)], \\
    \bmu_{2 k}^{(2)} &= \E[Z_0^2 Z_k^2 \mathbf{m}(Z_{-\infty:k}; \bbeta_0) - \sigma^2 Z_0^2 \mathbf{m}(Z_{-\infty:0}; \bbeta_0)].
\end{align*}
\begin{theorem}
    \label{thm-fin-acf-sq-asymp}
    Let $X_1, \ldots, X_n$ be observations from the model \eqref{eq-model} with $\E[Z_0^4] < \infty$ and $\hat{\bbeta}_n$ be an estimator of $\bbeta$ satisfying Assumptions~\ref{cond-m-est}, \ref{cond-l} and \ref{cond-tilde-z-2}.
    For a sample splitting sequence $\{(f_n, l_n)\}_{n \geq 1}$, write $\hat{Z}_j = \hat{Z}_j(\hat{\bbeta}_{f_n})\ (j = n - l_n + 1, \ldots, n)$.
    Then for all $h > 0$
    \begin{align*}
        \sqrt{l_n} (\hat{\rho}_{l_n}^{\hat{Z}^2}(1), \ldots, \hat{\rho}_{l_n}^{\hat{Z}^2}(h)) \gid \mathcal{N}_h (\mathbf{0},  \tilde{\Sigma}_{\gamma} ).
    \end{align*}
\end{theorem}


\begin{corollary}
    \label{thm-fin-acf-sq-asymp-cond}
    Assume the conditions of Theorem~\ref{thm-fin-acf-sq-asymp} and let $f_n = n / 2$ and $l_n = n$.
    Then for all $h > 0$, whenever $\tilde{M}_2 = - \tilde{M}_1 \Var(\mathbf{Q}) $ we have
    \begin{align*}
        \sqrt{n} \big( \hat{\rho}^{\hat{Z}^2}_{n}(1), \ldots, \hat{\rho}^{\hat{Z}^2}_{n}(h) \big)^\T \gid \mathcal{N}_h(\mathbf{0}, I_h).
    \end{align*}
\end{corollary}


\subsection{GARCH model}
\label{sec-acf-garch}

The GARCH($p, q$) model is given by
\begin{align*}
    X_j = \sigma_j Z_j,
\end{align*}
where the $Z_j$'s are iid random variables with mean 0 and variance 1 and the conditional variance is given by
\begin{align}
    \label{eq-garch-cond_var}
    \sigma_j^2 = \omega + \sum_{k = 1}^p \alpha_k X_{j - k}^2 + \sum_{\ell = 1}^{q} \beta_{\ell} \sigma_{j - \ell}^2,
\end{align}
where $\omega > 0, \alpha_k \geq 0, \beta_{\ell} \geq 0$.
Denote the parameter vector by $\btheta = (\omega, \alpha_1, \ldots, \alpha_p, \beta_1, \ldots, \beta_q)^\T$.
Iterating \eqref{eq-garch-cond_var} gives
\begin{align*}
    \sigma_j^2(\btheta) = c_0(\btheta) + \sum_{k = 1}^{\infty} c_k(\btheta) X_{j-k}^2,
\end{align*}
for some suitably defined functions $c_j$; see \citet{Berkesetal2003b, DavisWan2020} for details.
Based on an estimator $\btheta^\dagger$, at each time point $j$, an estimator of $\sigma^2_j(\btheta)$ is
\begin{align*}
    \sigma_j^2(\btheta^\dagger) = c_0(\btheta^\dagger) + \sum_{k = 1}^{\infty} c_k(\btheta^\dagger) X_{j - k}^2.
\end{align*}
The unobserved residuals are $\tilde{Z}_j (\btheta^\dagger) = X_j / (\tilde{\sigma}_j (\btheta^\dagger)$.
The conditional variance can be approximated by the finite sample approximation
\begin{align*}
    \hat{\sigma}_j^2(\btheta^\dagger) = c_0 (\btheta^\dagger) + \sum_{k = 1}^{j - 1} c_k(\btheta^\dagger) X_{j - k}^2,
\end{align*}
and thus the estimated residuals are given by $\hat{Z}_j (\btheta^\dagger) = X_j / (\hat{\sigma}_j (\btheta^\dagger) )$.
Define the parameter space by
\begin{align*}
    \mathbf{\Theta} = \{\mathbf{u} = (s_0, s_1, \ldots, s_p, t_1, \ldots, t_q) : t_1 + \cdots + t_q \leq \rho_0, \underline{u} \leq \min(\mathbf{u}) \le \max(\mathbf{u}) \le \bar{u}\},
\end{align*}
for some $0 < \underline{u} < \bar{u}$, $0 < \rho_0 < 1$ and $q \underline{u} < \rho_0$, and assume the following conditions:

\begin{assumption}
    \label{cond-g1}
    The true value $\btheta_0$ lies in the interior of $\mathbf{\Theta}$.
\end{assumption}

\begin{assumption}
    \label{cond-g2}
    For some $\zeta>0$,
    \begin{align*}
        \lim_{x\to0} x^{-\zeta} \mathbb{P} (|Z_0| \leq x) = 0.
    \end{align*}
\end{assumption}

\begin{assumption}
    \label{cond-g3}
    For some $\delta>0$,
    \begin{align*}
        \E|Z_0|^{4+\delta} < \infty.
    \end{align*}
\end{assumption}

\begin{assumption}
    \label{cond-g4}
    The GARCH($p,q$) representation is minimal, i.e., the polynomials $A(z)= \sum_{i=1}^p\alpha_iz^i$ and $B(z)=1-\sum_{j=1}^p\beta_jz^j$ do not have common roots.
\end{assumption}

\medskip
The quasi-maximum likelihood estimator for $\btheta_0$ \citep{Berkesetal2003b} is given by
\begin{align*}
    \hat{\btheta}_n := \argmax_{\mathbf{u} \in \mathbf{\Theta}} \sum_{j = 1}^{n} l_j (\mathbf{u}),
\end{align*}
where
\begin{align*}
    l_j(\mathbf{u}) := -\frac12 \log \hat{\sigma_j}^2 (\mathbf{u}) - \frac{X_j^2}{2\hat{\sigma}_j^2(\mathbf{u})}.
\end{align*}
Provided that Assumptions \ref{cond-g1}--\ref{cond-g4} are satisfied, the quasi maximum likelihood estimator $\hat{\btheta}_n$ is consistent and asymptotically normal.

Assumptions \ref{cond-m-est}--\ref{cond-tilde-z-2} are verified in Section~\ref{app-garch-assump}.
Due to Theorem~\ref{thm-fin-acf-sq-asymp}, we have
\begin{align*}
    \mathbf{Q} &\sim \mathcal{N}_p (\mathbf{0}, \mathbf{B}_0^{-1} \mathbf{A}_0 \mathbf{B}_0^{-1}), \\
    \mathbf{m}(Z_{-\infty:j}; \btheta_0) &= \frac{1}{2} (1 - Z_j^2) \mathbf{B}_0^{-1} \frac{\partial}{\partial \btheta} \log \sigma_j^2 (\btheta_0), \\
    \mathbf{L}_j(\btheta_0) &= -\frac{1}{2} Z_j \frac{\partial}{\partial \btheta}  \log \sigma_j^2 (\btheta_0),
\end{align*}
where $\mathbf{A}_0 = \Cov\big[ \frac{\partial}{\partial \btheta} l_0(\btheta_0) \big]$, $\mathbf{B}_0 = \E \big[ \frac{\partial^2}{\partial \btheta^2} l_0(\btheta_0) \big]$.
Recall that $\sigma^2 = 1$ and using $\frac{\partial}{\partial \btheta} \log \sigma_j^2(\btheta_0)$ is independent of $Z_j$ we have
\begin{align*}
    \bmu_{1 j}^{(2)} &= - \E \Big[ Z_0^2 Z_j^2 \frac{\partial}{\partial \btheta} \log \sigma_j^2(\btheta_0) - Z_0^2 \frac{\partial}{\partial \btheta} \log \sigma_0^2(\btheta_0) \Big] \\
    &= - \E \Big[ (Z_{0}^2 - 1) \frac{\partial}{\partial \btheta} \log \sigma_j^2(\btheta_0) \Big], \\
    \bmu_{2j}^{(2)} &= \frac{1}{2} \mathbf{B}_0^{-1} \E \Big[ Z_0^2 Z_j^2 (1 - Z_j^2)  \frac{\partial}{\partial \btheta} \log \sigma_j^2(\btheta_0) - Z_0^2 (1 - Z_0^2)  \frac{\partial}{\partial \btheta} \log \sigma_0^2(\btheta_0) \Big] \\
    &= - \frac{1}{2} (\E[Z_0^4] - 1)  \mathbf{B}_0^{-1} \E \Big[ (Z_{0}^2 - 1)  \frac{\partial}{\partial \btheta} \log \sigma_j^2(\btheta_0) \Big].
\end{align*}
From Remark 4.5 of \citet{Berkesetal2003b}, $\mathbf{A}_0 = -\frac{1}{2} (\E[Z_0^4] - 1) \mathbf{B}_0$ and thus the condition $\bmu_{2 j}^{(2)} = -  \Var(\mathbf{Q}) \bmu_{1 j}^{(2)}$ holds for $1 \leq j \leq h$.
Thus Corollary~\ref{thm-fin-acf-sq-asymp-cond} implies
\begin{align*}
    \sqrt{n} (\hat{\rho}_{n}^{\hat{Z}^2}(1), \ldots, \hat{\rho}_{n}^{\hat{Z}^2}(h)) \gid \mathcal{N}_h(\mathbf{0}, I_h).
\end{align*}
With $f_n = l_n = n$ we recover Theorem 2.2 of \citet{Berkesetal2003a}, with the limiting covariance matrix $I_h - \tilde{M}_1 \Var(\mathbf{Q}) \tilde{M}_1^\T / (\E[Z_0^4] - 1)^2$.

\begin{figure}[t]
    \centering
    \includegraphics[width=0.8\textwidth]{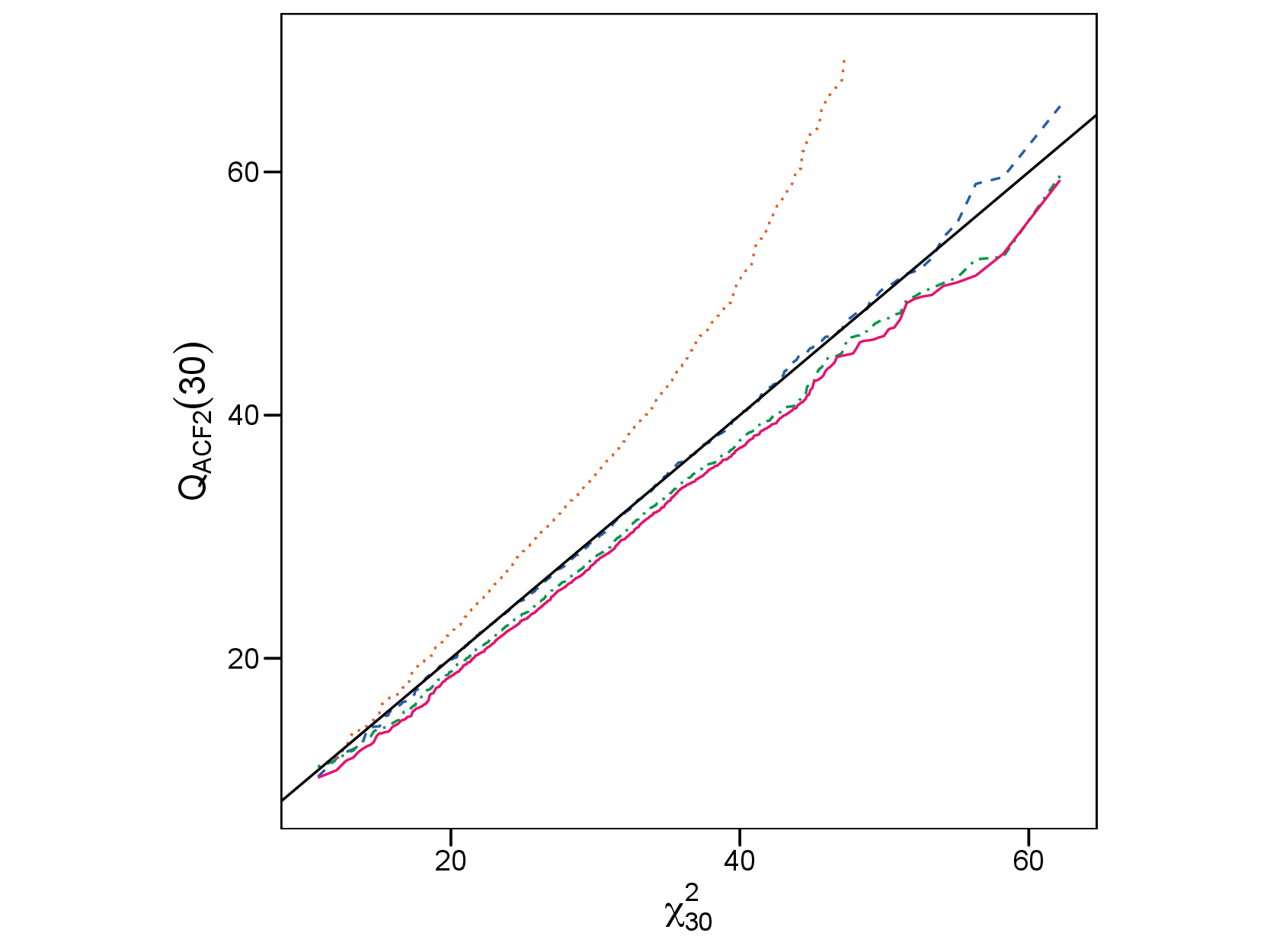}
    \caption{
        Quantile-quantile plot of $Q_{ACF2}(30)$ versus $\chi^2_{30}$ for $f_n$ being 2000 (orange, dotted), 5000 (blue, dashed), 8000 (green, dot-dash) and 10000 (red, solid), and $l_n = n = 10000$.
        The empirical distribution of $Q_{ACF2}(30)$ is based on 1000 simulations.
    }
    \label{fig-garch11-normal-acf_sq-qq}
\end{figure}

We demonstrate this result for a GARCH$(1, 1)$ model.
In each simulation, a time series of length $n = 10000$ is generated from a GARCH$(1, 1)$ model with parameter
\begin{align*}
    \btheta_0 = (\omega, \alpha, \beta) = (0.5, 0.1, 0.8),
\end{align*}
where the innovations are standard Gaussian.
To illustrate the results of sample splitting, four different splits are considered wherein $l_n = 10000$ and $f_n$ is one of 2000, 5000, 8000, 10000.
For each simulated time series, we calculate the residual ACF for each sample split at lags 1 to 30 and calculate the sum of squares
\begin{align*}
    Q_{ACF2}(30) = n \sum_{j = 1}^{30} \Big( \hat{\rho}^{\hat{Z}^2}_{n} (j) \Big)^2.
\end{align*}
Figure~\ref{fig-garch11-normal-acf_sq-qq} displays the qqplot against the reference distribution of $\chi^2_{30}$ for each split.
As predicted under Corollary~\ref{thm-fin-acf-sq-asymp-cond}, at $f_n = 5000$ the qqplot closely follows the theoretical quantiles.
Furthermore, the distribution with $f_n = 2000$ is stochastically larger than $\chi^2_{30}$ whereas the distributions with $f_n = 8000, 10000$ are stochastically smaller.

\section{Asymptotics for the ADCF}
\label{sec-adcf}


The ACF based tests do not detect nonlinear dependencies among residuals.
Further, the ACF being zero does not imply there is no dependence among the residuals.
In this section, we focus on the auto-distance correlation function (ADCF) which is a measure that tests independent residuals rather than uncorrelated residuals.
We first introduce the distance covariance as a measure of independence between two random vectors before applying this to testing independence of lagged residuals.

For a random vector $(\mathbf{X}, \mathbf{Y}) \in \R^{p + q}$, distance covariance computes the integral squared distance between the characteristic functions
\begin{align*}
    T(\mathbf{X}, \mathbf{Y}; \mu) := \int_{(s, t) \in \R^{p + q}} |\varphi_{\mathbf{X}, \mathbf{Y}}(s,t) - \varphi_{\mathbf{X}}(s) \varphi_{\mathbf{Y}}(t)|^2 d\mu (s, t),
\end{align*}
where $\varphi_{\mathbf{X}, \mathbf{Y}}(s, t)$ is the characteristic function of $(\mathbf{X}, \mathbf{Y})$, $\varphi_{\mathbf{X}}(s) = \varphi_{\mathbf{X}, \mathbf{Y}}(s, 0)$, $\varphi_{\mathbf{Y}}(t) = \varphi_{\mathbf{X}, \mathbf{Y}}(0, t)$, and $\mu$ is some weight function.
Choosing a finite measure for the weight function guarantees the finiteness of $T$.
In \citet{Szekelyetal2007} the weight function used is the infinite measure
\begin{align*}
    d\mu_s(s, t) = {ds dt}/{(c_p c_q |s|^{p + \alpha} |t|^{q + \alpha})}, \;\; s \in \R^p,\; t \in \R^q,
\end{align*}
where $0 < \alpha < 2$, and $c_p, c_q \in \R$ are constants such that
\begin{align*}
    \int (1 - \cos(s^\T x)) c_d |s|^{-s-\alpha} ds = |x|^{\alpha},
\end{align*}
for any $d \geq 1$.
With this choice of weight function, one obtains
\begin{align*}
    T(\mathbf{X}, \mathbf{Y}; \mu_s)
    &= \E[ |X - \dot{X}|^{\alpha} |Y - \dot{Y}|^{\alpha}]
    + \E[ |X - \dot{X}|^{\alpha}] \E[|Y - \dot{Y}|^{\alpha}] \nonumber \\
    &- 2 \E[ |X - \dot{X}|^{\alpha} |Y - \ddot{Y}|^{\alpha}].
\end{align*}
See Section 2 of \citet{Davisetal2018} for further discussion and details.

Consider now the auto-distance covariance at lag $h$ of the sample splitting based residuals, that is the distance-covariance of the $h$-lagged sample
\begin{align*}
    \hat{T}^{\hat Z}_{l_n}(h) &=
    \frac{1}{(l_n - h)^2} \sum_{j, k = n - l_n + 1}^{n - h} \hat{\mu}(\hat{Z}_{j} - \hat{Z}_{k}, \hat{Z}_{j + h} - \hat{Z}_{k + h}) \nonumber\\
    &+ \frac{1}{(l_n - h)^4} \sum_{j, k, \ell, m = n - l_n + 1}^{n - h} \hat{\mu}(\hat{Z}_{j} - \hat{Z}_{k}, \hat{Z}_{\ell + h} - \hat{Z}_{m + h}) \nonumber\\
    &- \frac{2}{(l_n - h)^3} \sum_{j,k, \ell = n - l_n + 1}^{n - h} \hat{\mu}(\hat{Z}_{j} - \hat{Z}_{k}, \hat{Z}_{j + h} - \hat{Z}_{\ell + h}),
\end{align*}
where $\hat{\mu}(x, y) = \int \exp(isx + ity) d\mu(s, t)$ is the Fourier transform of the weight measure $\mu$.

The ADCF is then
\begin{align*}
    \hat{R}^{\hat{Z}}_{l_n}(h) = \hat{T}^{\hat{Z}}_{l_n}(h) / \hat{T}^{\hat{Z}}_{l_n}(0).
\end{align*}
For the true noise, we have $\hat{T}_{l_n}^{Z}(h) \gip T^{Z}(h)$, where $T^{Z}(0) > 0$ and $T^{Z}(h) = 0$ for $h \geq 1$; see \citet{Davisetal2018}.

The result for sample splitting using the ADCF closely mirrors the result for the ACF.
In this case the convergence to the Gaussian process of the lagged characteristic functions determine the asymptotics.
Similar to \eqref{eq-fin-acf-var}, the variance of the asymptotic distribution of the ADCF depends on the sample splitting sequences.
It follows from Theorem 3.2 of \citet{Davisetal2018} that for the true noise we have
\begin{align*}
    l_n \hat{R}_{l_n}^{Z}(h) \gid \frac{\| G_h \|_{\mu}^2}{T^{Z}(0)},
\end{align*}
where $G_h$ is a Gaussian process with covariance function
\begin{align}
    \label{eq-fin-dcf-g-cov}
    \sigma_T((s_1, t_1), (s_2, t_2)) = [\varphi_Z(s_1 - s_2) - \varphi_Z(s_1)\overline{\varphi_Z(s_2)}] [\varphi_Z(t_1 - t_2) - \varphi_Z(t_1) {\overline{\varphi_Z(t_2)}}],
\end{align}
and
\begin{align*}
    \| G_h \|_{\mu}^2 = \int_{\R^2} |G_h (s, t)|^2 d\mu(s, t).
\end{align*}
For the sample splitting based residuals however, $\hat{T}_{l_n}^{\hat{Z}}(h)$ converges to $\| G_h + \xi_h \|_{\mu}^2$, the squared integral of the Gaussian process $G_h + \xi_h$ which has covariance function
\begin{align}
    \label{eq-fin-dcf-cov}
    \Cov \big( (G_h + \xi_h)(s_1, t_1), (G_h + \xi_h)(s_2, t_2) \big)
    &= \sigma_T((s_1, t_1), (s_2, t_2)) \nonumber \\
    &+ k_{ov} \bnu_{1 h}^\T (s_1, t_1) \overline{\bnu_{2 h} (s_2, t_2)} \nonumber \\
    &+ k_{ov} \bnu_{2 h}^\T (s_1, t_1) \overline{\bnu_{1 h} (s_2, t_2)} \nonumber \\
    &+ k_{ra} \bnu_{1 h}^\T (s_1, t_1) \Var(\mathbf{Q}) \overline{\bnu_{1 h} (s_2, t_2)},
\end{align}
where
\begin{align*}
    \bnu_{1 h}(s, t) &= it \E [(e^{i s Z_0} - \varphi_Z(s)) e^{i t Z_h} \mathbf{L}_h (\bbeta_0)], \\
    \bnu_{2 h}(s, t) &= \E [(e^{i s Z_0} - \varphi_Z(s)) e^{i t Z_h} \mathbf{m}(Z_{-\infty:h}; \bbeta_0)].
\end{align*}
We now state the main result.
\begin{theorem}
    \label{thm-fin-dcf-asymp}
    Let $X_1, \ldots, X_n$ be observations from the model \eqref{eq-model} and $\hat{\bbeta}_n$ be an estimator of $\bbeta$ satisfying Assumptions~\ref{cond-m-est}--\ref{cond-tilde-z}.
    For a sample splitting sequence $\{(f_n, l_n)\}_{n \geq 1}$, write $\hat{Z}_j = \hat{Z}_j(\hat{\bbeta}_{f_n})\ (j = n - l_n + 1, \ldots, n)$.
    If the weight function $\mu$ satisfies
    \begin{align}
        \label{eq-weight}
        \int_{\mathbb{R}^2} \big[ (1 \wedge |s|^2) (1 \wedge |t|^2) +  (s^2 + t^2) \mathbb{I}(|s| \wedge |t|>1) \big] d\mu(s, t) < \infty,
    \end{align}
    then for all $h > 0$
    \begin{align*}
        l_n \hat{R}_{l_n}^{\hat{Z}}(h) \gid \frac{\|G_h + \xi_h\|^2_{\mu}}{T^{Z}(0)}.
    \end{align*}
\end{theorem}

From the form of the covariance in \eqref{eq-fin-dcf-cov} we see exactly how the model and the estimation procedure influence the ADCF of the residuals.
Sample splitting provides the opportunity to choose settings where upon the covariance in \eqref{eq-fin-dcf-cov} is \eqref{eq-fin-dcf-g-cov}, the covariance of $G_h$.
We show in the following that ARMA and GARCH models satisfy $\bnu_{2 h} (s, t) = - \tau_Z(t) \Var(\mathbf{Q}) \bnu_{1 h} (s, t)$, where $\tau_Z(\cdot)$ is some real valued function depending only on the distribution of $Z$.
In this case the covariance simplifies to
\begin{align}
    \label{eq-fin-dcf-cov-cond}
    \Cov \big( (G_h + \xi_h) (s_1, t_1)&, (G_h + \xi_h) (s_2, t_2) \big) 
    = \sigma_T((s_1, t_1), (s_2, t_2)) \nonumber \\
    &+ \big(k_{ra} - (\tau_Z(t_1) + \tau_Z(t_2))k_{ov} \big) \bnu_{1 h}^\T (s_1, t_1) \Var(\mathbf{Q}) \overline{\bnu_{1 h} (s_2, t_2)},
\end{align}
Furthermore when the noise has a Gaussian distribution, in both the ARMA and GARCH cases $\tau_Z$ is the constant function.
We thus have the following result.

\begin{corollary}
    \label{thm-fin-dcf-cond}
    Assume the conditions of Theorem~\ref{thm-fin-dcf-asymp} and that $\bnu_{2 h} (s, t) = - \Var(\mathbf{Q}) \bnu_{1 h} (s, t)$.
    Then with $f_n = n / 2$ and $l_n = n$, we have
    \begin{align*}
        n \hat{R}_{n}^{\hat Z}(h) \gid \frac{\|G_h\|_{\mu}^2}{T^Z(0)}.
    \end{align*}
\end{corollary}

The residual ADCF values with $f_n = n / 2$ and $l_n = n$ can thus be directly compared to the quantiles of the distribution of the ADCF of true noise.
Assuming that this noise is Gaussian the 95\% confidence quantile is obtained via simulation.
For example if $\mu = \mu_1 \times \mu_2$, where $\mu_i$ is the $\mathcal{N}(0, 0.5)$ distribution, the 95\% quantile equals $4.5 / n$.
Note however that $\mu_s$ does not satisfy \eqref{eq-weight}.
Furthermore, the sign of $k_{ra} - 2k_{ov}$ can be used to determine the stochastic order between the residual ADCF and true noise ADCF.
We employ $\preceq$ to denote stochastic order and write $U \preceq V$ to mean that $U$ is stochastically dominated by $V$, where $U, V$ are random variables.
To compare the stochastic ordering among $G_h + \xi_h$ and $G_h$ we need the following result.

\medskip

\begin{lemma}
    \label{thm-stoch}
    Let $G_h + \xi_h$ be a mean zero Gaussian process with covariance function \eqref{eq-fin-dcf-cov-cond}.
    If $\tau_Z(t) \leq 1$ for all $t$ and $k_{ra} \geq 2k_{ov}$ then $\|G_h + \xi_h\|^2_{\mu} \succeq \|G_h\|^2_{\mu}$.
    Similarly, if $\tau_Z(t) \geq 1$ for all $t$ and $k_{ra} \leq 2k_{ov}$ then $\|G_h + \xi_h\|^2_{\mu} \preceq \|G_h\|^2_{\mu}$.
\end{lemma}

\medskip
Suppose $\bnu_{2 h} (s, t) = - \Var(\mathbf{Q}) \bnu_{1 h} (s, t)$ holds so thats the covariance of $G_h + \xi_h$ is given by \eqref{eq-fin-dcf-cov-cond}.
For the full model $f_n = l_n = n$, the residual ADCF is stochastically smaller than the true noise ADCF.
Making $f_n$ smaller than $n / 2$ with $l_n = n$ makes $k_{ra} - 2k_{ov}$ positive so that the residual ADCF is stochastically larger than the true noise ADCF.
As Corollary~\ref{thm-fin-dcf-cond} shows, $k_{ra} - 2 k_{ov} = 0$ is the regime where independent residuals are obtained.
Similar to the ACF case, employing disjoint splits is not helpful.
In fact for a general model with $k_{ov} = 0$ in \eqref{eq-fin-dcf-cov}, Lemma~\ref{thm-stoch} shows that the ADCF of the residuals will be stochastically larger than the ADCF of the true noise at each lag.

\begin{remark}
    Note that sample splitting does not make the correction term $\xi_h$ vanish - this is always present and it obfuscates the limiting properties of the ADCF.
    One can deduce from \eqref{eq-fin-acf-var} that $\xi_h (s, t) = \sqrt{k_{ra}} \bnu_{1 h}^\T (s, t) \mathbf{Q}$ and $\Cov(G_h (s, t), \mathbf{Q}) = k_{ov} \bnu_{2 h}^\T (s, t) / \sqrt{k_{ra}}$; so $\xi_h$ is simply a scaling of $\mathbf{Q}$ that depends on $\bnu_{1 h}$.
    The trick, so-to-speak, is to leverage this dependence and use sample splitting settings wherein $G_h + \xi_h \eqd G_h$.
    The fact that this is achievable simply by adjusting the sample splitting sequences is the strength of this method.
\end{remark}

\begin{figure}[t]
    \centering
    \begin{subfigure}[b]{0.45\textwidth}
        \centering
        \includegraphics[width=\textwidth]{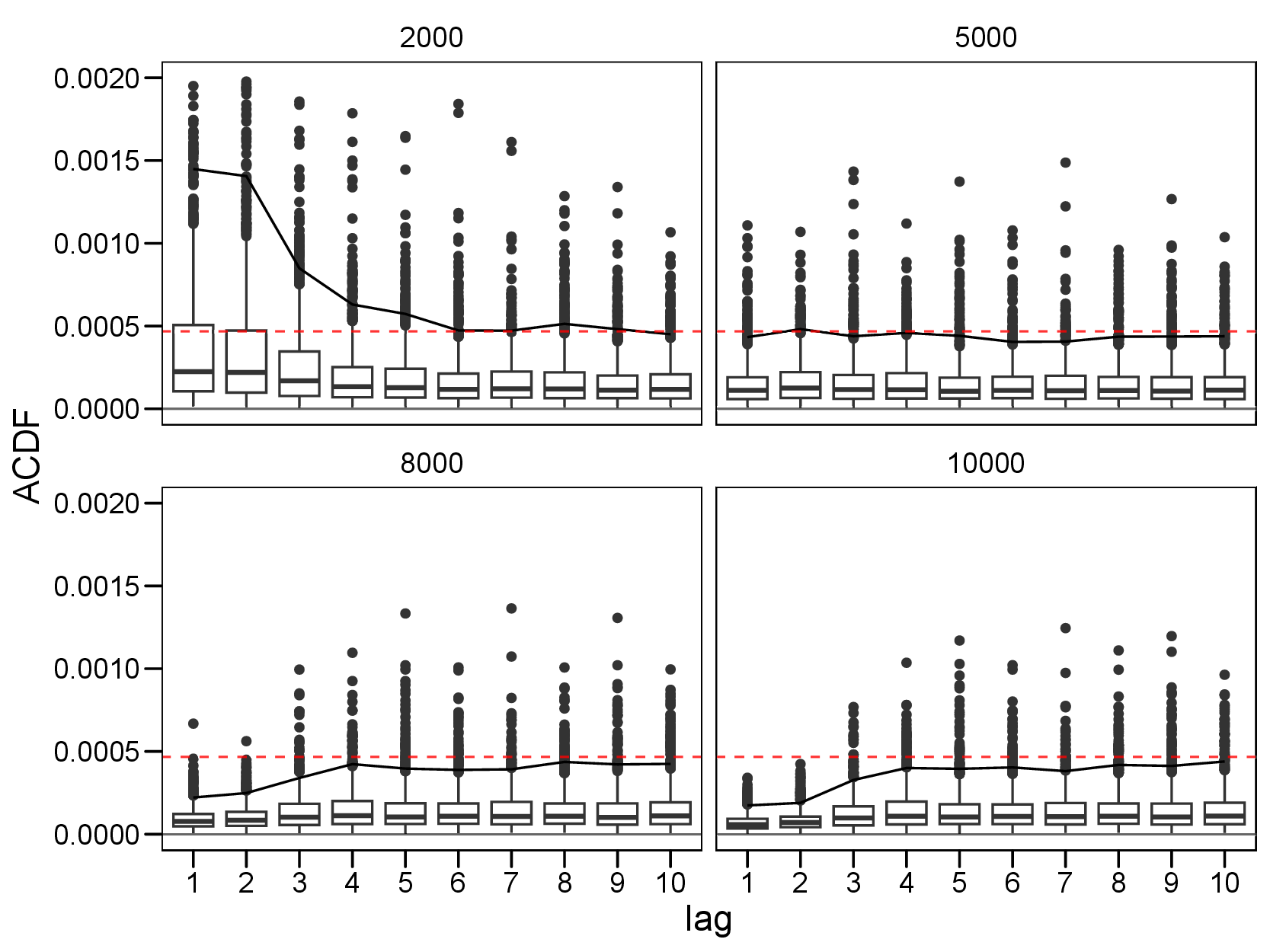}
        \caption{
            Gaussian
        }
        \label{fig-arma-normal-dcf-gauss}
    \end{subfigure}
    \hfill
    \begin{subfigure}[b]{0.45\textwidth}
        \centering
        \includegraphics[width=\textwidth]{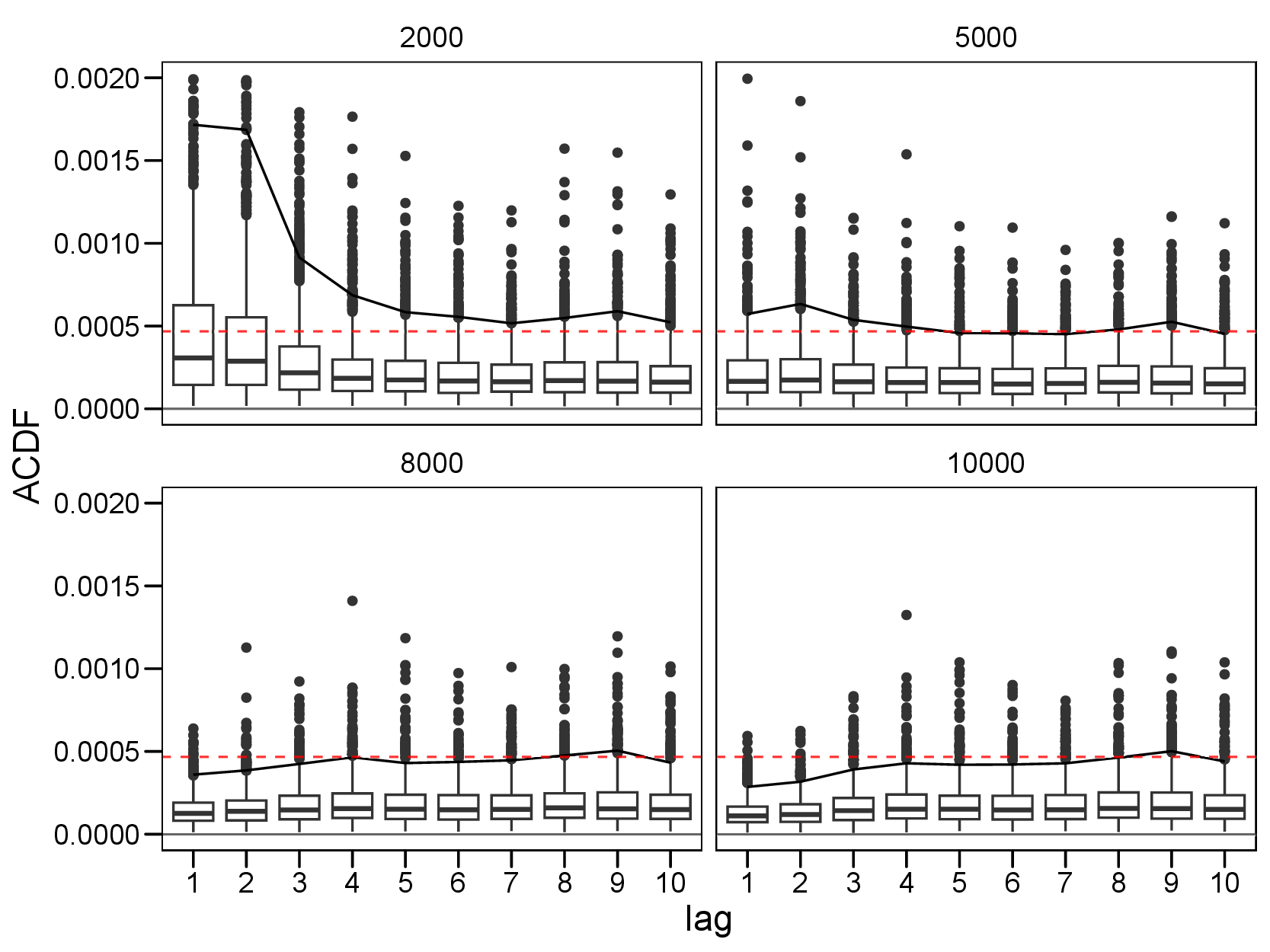}
        \caption{
            Laplace
        }
        \label{fig-arma-laplace-dcf-gauss}
    \end{subfigure}
    \caption{
        Simulated 95\% quantile of ADCF at lags 1 to 10 for an ARMA(2, 1) model with (a) Gaussian and (b) Laplace noise.
        Here $l_n$ is 10000 and each subplot corresponds to the value of $f_n$.
        The quantiles are obtained from 1000 simulation of the AR process, each of length 10000.
        The dashed red lines denote the 95\% quantile assuming Gaussian noise, that is $4.41 / n$.
    }
    \label{fig-ar10-dcf}
\end{figure}

\begin{example}[ARMA model]
Consider the ARMA model setting from Section~\ref{sec-acf-arma}.
We have already verified all the hypotheses of Theorem~\ref{thm-fin-dcf-asymp}.
We claim that $\bnu_{2 h}(s, t) = -\tau_Z(t)  \Var(\mathbf{Q}) \bnu_{1 h}(s, t)$, where $\tau_Z(t) = -\varphi_Z'(t)/(t\sigma^2\varphi_Z(t))$.
It is clear that $\tau_Z(t) = 1$ for all $t$ holds if and only if $Z$ is Gaussian with mean zero.
In this case we have Corollary~\ref{thm-fin-dcf-cond} and thus
\begin{align*}
    n \hat{R}_{n}^{\hat Z}(h) \gid \frac{\|G_h\|_{\mu}^2}{T^Z(0)}.
\end{align*}
We have $\bnu_{1 h}(s, t) = it \varphi_Z(t) \E[(e^{isZ_0} - \varphi_Z(s)) \mathbf{L}_h(\bbeta_0)]$ due to independence of $Z_h$ and $\mathbf{L}_h(\bbeta_0)$.
Then
\begin{align*}
    \bnu_{2 h}(s, t) = - \Var(\mathbf{Q}) \E[(e^{isZ_0} - \varphi_Z(s)) e^{itZ_h} \mathbf{L}_h(\bbeta_0) Z_h] / \sigma^2 = - \tau_Z(t) \Var(\mathbf{Q}) \bnu_{1 h}(s, t),
\end{align*}
where $\tau_Z(t) = \E[Z_h e^{itZ_h}]/ (i \sigma^2 t \varphi_Z(t)) =  -\varphi_Z'(t)/(\sigma^2 t \varphi_Z(t))$.
For the case where $Z$ is not Gaussian, one needs to establish upper (or lower) bounds on $\tau_Z(t)$.
\begin{lemma}
    \label{thm-chfn-ineq}
    If $Z$ has Laplace or Student's t distribution with mean zero, then for all $t \in \R$
    \begin{align}
        \label{eq-chfn-ineq}
        \frac{-\varphi_Z'(t)}{t \sigma^2 \varphi_Z(t)} \leq 1.
    \end{align}
\end{lemma}
The Laplace case follows easily but for the Student's t, one requires a Turan type inequality for modified Bessel functions; see Section~\ref{app-chfn-ineq} for the proof.
For such a $Z$ we can conclude using Lemma~\ref{thm-stoch} that whenever $k_{ra} \geq 2k_{ov}$, the residual ADCFs are stochastically larger than the true noise ADCFs.
A future direction of this work would be in proving \eqref{eq-chfn-ineq} for a larger set of distributions of the noise.

To demonstrate these results, consider simulating from the same ARMA(2, 1) model as in Section~\ref{sec-acf-arma}.
The ADCF is calculated with the weight measure $\mu = \mu_1 \times \mu_2$, where $\mu_i$ is the $\mathcal{N}(0, 0.5)$ which satisfies \eqref{eq-weight}.
Figure~\ref{fig-arma-normal-dcf-gauss} displays the simulated asymptotic distribution of the residual ADCF across four sample splits when the distribution of the noise is Gaussian.
At $f_n = 5000$ the ADCF corresponds exactly to the theoretical Gaussian quantile as is expected due to Corollary~\ref{thm-fin-dcf-cond}.
When the noise has Laplace distribution with mean zero and variance one, the quantiles at $f_n = 5000$ do not correspond to iid residuals as can be observed in Figure~\ref{fig-arma-laplace-dcf-gauss}.
\end{example}

\begin{example}[GARCH model]
In the GARCH setting we have
\begin{align*}
    \bnu_{1 h}(s, t)
    &= - \frac{1}{2} i t \E \Big[(e^{i s Z_0} - \varphi_Z(s)) Z_h e^{i t Z_h}  \frac{\partial}{\partial \btheta}  \log \sigma_h^2 (\btheta_0)\Big], \\
    \bnu_{2 h}(s, t)
    &= \frac{1}{2} \mathbf{B}_0^{-1} \E \Big[ (e^{i s Z_0} - \varphi_Z(s)) e^{i t Z_h} (1 - Z_h^2) \frac{\partial}{\partial \btheta} \log \sigma_h^2 (\btheta_0) \Big].
\end{align*}
Due to independence of $Z_h$ and $\frac{\partial}{\partial \btheta} \log \sigma_h^2(\btheta_0)$, $\E[Z_h e^{itZ_h}] = -i \varphi_Z'(t)$, $\E[Z_h^2 e^{itZ_h}] = - \varphi_Z''(t)$ and $\mathbf{A}_0 = -\frac{1}{2} (\E[Z_0^4] - 1) \mathbf{B}_0$, we get $\bnu_{2 h}(s, t) = - \tau_Z(t) \Var(\mathbf{Q}) \bnu_{1 h}(s, t)$, where
\begin{align*}
    \tau_Z(t) = \frac{-2}{\E[Z^4] - 1} \cdot \frac{\varphi_Z(t) + \varphi_Z''(t)}{t \varphi_Z'(t)}.
\end{align*}
If $Z$ is Gaussian then $\lambda_Z(t) = 1$ for all $t$.
In this case we can apply Corollary~\ref{thm-fin-dcf-cond}.
If $Z$ is Laplace with mean zero and variance one, then $\tau_Z(t) \leq 1$.
Thus whenever $k_{ra} \geq 2k_{ov}$, Lemma~\ref{thm-stoch} implies the residual ADCFs will be stochastically larger than true noise ADCFs.
\end{example}

\section{A Power Calculation}

\begin{table}[t]
    \def~{\hphantom{0}}
    \caption{
            Rejection rates (in percentage) for the goodness-of-fit of AR(1) model on data generated from AR(1) model with GARCH(1, 1) noise for $n = 500$.
            The rates are based on 1000 simulations and are calculated for varying number of lags $h$, AR(1) coefficient $\phi$ and GARCH coefficients $(\alpha, \beta)$.
        }
    { \small
    \begin{tabular}{ccccccccccc}
            Test & $h$ & \multicolumn{3}{l}{$\phi = 0.2$} & \multicolumn{3}{l}{$\phi = 0.5$} & \multicolumn{3}{l}{$\phi = 0.8$} \\
            & & \multicolumn{3}{l}{$(\alpha, \beta)$} & \multicolumn{3}{l}{$(\alpha, \beta)$} & \multicolumn{3}{l}{$(\alpha, \beta)$} \\
            & & (0, 0) & (.3, .3) & (.3, .6) & (0, 0) & (.3, .3) & (.3, .6) & (0, 0) & (.3, .3) & (.3, .6) \\
            $Q_{ACF}$ & 2 & 4.7 & 14.0 & 24.8 & 4.9 & 18.0 & 24.7 & 7.9 & 15.8 & 28.7 \\ 
            & 5 & 5.8 & 15.3 & 33.2 & 5.1 & 17.7 & 34.2 & 6.8 & 14.9 & 38.1 \\  
            & 8 & 5.4 & 12.0 & 35.7 & 4.5 & 14.2 & 34.1 & 5.4 & 14.4 & 40.5 \\ 
            $Q_{ADCF}$ & 2 & 3.8 & 71.2 & 98.9 & 3.0 & 73.5 & 98.2 & 4.5 & 71.3 & 98.2 \\ 
            & 5 & 5.1 & 73.9 & 99.9 & 3.9 & 73.2 & 99.9 & 5.0 & 72.0 & 99.9 \\ 
            & 8 & 5.4 & 74.4 & 100.0 & 3.9 & 75.6 & 100.0 & 6.5 & 77.4 & 100.0 \\ 
            $Q_{LB}$ & 2 & 5.1 & 9.3 & 19.2 & 4.7 & 11.8 & 19.2 & 7.7 & 18.6 & 28.9 \\ 
            & 5 & 5.1 & 11.3 & 29.9 & 4.7 & 13.4 & 30.1 & 5.1 & 12.6 & 36.0 \\ 
            & 8 & 5.8 & 10.3 & 31.7 & 4.4 & 10.1 & 31.5 & 4.4 & 11.0 & 38.5
        \end{tabular}
    }
    \label{tab-power-garch}
\end{table}

We empirically estimate power against an alternative with model miss-specification.
The data is generated from an AR(1) model where the noise is uncorrelated but not independent.
We generate the noise $Z_t$  from a GARCH(1, 1) model $Z_j = \sigma_j \varepsilon_j$, where $\varepsilon_j \sim \mathcal{N}(0, 1)$ are iid and
\begin{align*}
    \sigma_j^2 = 1 + \alpha Z_{j - 1}^2 + \beta \sigma_{j - 1}^2,
\end{align*}
for GARCH parameters $(\alpha, \beta)$ such that $\alpha, \beta \geq 0$, $\alpha + \beta < 1$.
Given the observations we would like to ensure that the goodness-of-fit tests based on the ACF and ADCF are effective at identifying lack of fit if the fitted model is not the truth.
\begin{align*}
    H_0: \text{the data is generated from an AR(1) model}; \\
    H_1: \text{the data is not generated from an AR(1) model}.
\end{align*}
In this case, we will compare the ACF and ADCF of the residuals based on sample splitting with $f_n = n / 2$ and $l_n = n$.
The test statistic for the ACF will be the chi-squared test
\begin{align*}
    Q_{ACF}(h) = n \sum_{k = 1}^{h} \big( \hat{\rho}_n^{\hat{Z}}(k) \big)^2
\end{align*}
for lags $h = 2, 5, 8$.
For the test using ADCF, we consider the sum of the ADCF values up to lag $h$ given by
\begin{align*}
    Q_{ADCF}(h) = n \sum_{k = 1}^{h} \hat{R}_{n}^{\hat{Z}}(k),
\end{align*}
and $H_0$ is rejected when this is above a critical value.
Here the critical value are obtained by simulations for Gaussian noise; in particular the values obtained were 7.84, 14.2 and 20.0 at lags 2, 5 and 8 respectively.

The standard method is the Ljung-Box test statistic \citep{LjungBox1978} and will be used as the benchmark.
The test statistic is based on the fitted residuals $\{\hat{Z}_j\}$ obtained under $f_n = l_n = n$ and is given by
\begin{align*}
    Q_{LB}(h) = n(n + 2) \sum_{k = 1}^{h} \frac{1}{n - k} \big( \hat{\rho}_n^{\hat{Z}}(k) \big)^2.
\end{align*}
We reject $Q_{LB}$ if it is larger than the critical value $\chi^2_{h - 1}(0.95)$.
The rejection rates based on 1000 replications are shown in Table \ref{tab-power-garch}.
The columns with $\alpha = \beta = 0$ correspond to the noise process being iid and thus the rejection rates correspond to the size of the test.
We note that all three tests control the type I error.
The ADCF test performs the best in this case with the hightest power.
The ACF test beats out the Ljung-Box test in all cases signifying no loss of power, even though both tests are only relying on the ACF values.


\section*{Acknowledgement}
The research of R. A. Davis was supported in part by NSF grant DMS 2310973.
We would like to thank Professor Miguel A. Delgado for bringing to our attention the paper \citet{Durbin1976} and the connections with our results.
We acknowledge computing resources from Columbia University's Shared Research Computing Facility project, which is supported by NIH Research Facility Improvement Grant 1G20RR030893-01, and associated funds from the New York State Empire State Development, Division of Science Technology and Innovation (NYSTAR) Contract C090171, both awarded April 15, 2010.



\appendix
\section{Proofs}

In this section, $c$ denotes some constant whose value may change from line to line in the proofs.

\subsection{Proof of Theorem~\ref{thm-fin-acf-asymp}}
\label{app-acf-asymp}

For notational convenience, we write $\hat{\bbeta} = \hat{\bbeta}_{f_n}$.
We will first consider the limiting properties of the ACF at lag $h$.
The multivariate case will follow from a straightforward extension of the proof.
Writing the autocovariance of the residuals in terms of the autocovariance of the noise process, we have
\begin{align}
    \label{eq-acf-re_exp}
    (l_n - h) \hat{\gamma}_{l_n}^{\hat{Z}}(h)
    &= \sum_{j = n - l_n + 1}^{n - h} Z_{j} Z_{j + h} \nonumber \\
    &+ \sum_{j = n - l_n + 1}^{n - h} Z_{j}(\hat{Z}_{j + h} - Z_{j + h}) \nonumber \\
    &+ \sum_{j = n - l_n + 1}^{n - h} (\hat{Z}_{j} - Z_{j})Z_{j + h} \nonumber \\
    &+ \sum_{j = n - l_n + 1}^{n - h} (\hat{Z}_{j} - Z_{j})(\hat{Z}_{j  + h} - Z_{j + h}).
\end{align}
For the first term, the central limit theorem yields $\sqrt{l_n} \hat{\gamma}_{l_n}^{Z}(h) \gid \mathcal{N}(0, \sigma^4)$. 
Due to Cauchy-Schwarz, Assumption~\ref{cond-tilde-z} and $\Var(Z_0) < \infty$, we have as $n \rightarrow \infty$
\begin{align*}
    \frac{1}{\sqrt{l_n}} \sum_{j = n - l_n + 1}^{n - h} Z_{j} (\hat{Z}_{j + h} - \tilde{Z}_{j + h}) \gip 0,
\end{align*}
so that for the second term in \eqref{eq-acf-re_exp}
\begin{align*}
    \sum_{j = n - l_n + 1}^{n - h} Z_{j} (\hat{Z}_{j  + h} - Z_{j + h})
    = \sum_{j = n - l_n + 1}^{n - h} Z_{j} (\tilde{Z}_{j + h} - Z_{j + h}) + o_P(1 / \sqrt{l_n}).
\end{align*}
A Taylor expansion using Assumption~\ref{cond-l} yields
\begin{align*}
    \frac{1}{\sqrt{l_n}} \sum_{j = n - l_n + 1}^{n - h} Z_{j} (\tilde{Z}_{j + h} - Z_{j + h})
    = \sqrt{l_n} (\hat{\bbeta} - \bbeta_0)^\T \cdot \frac{1}{l_n}\sum_{j = n - l_n + 1}^{n - h} Z_{j} \mathbf{L}_{j + h}(\bbeta^*)
\end{align*}
where $\bbeta^* = \bbeta_0 + \varepsilon(\hat{\bbeta} - \bbeta_0)$ for some $\varepsilon \in [0, 1]$.
If we show
\begin{align}
    \label{eq-fin-acf-z-l}
    \frac{1}{l_n} \sum_{j = n - l_n + 1}^{n - h} | Z_j | \| \mathbf{L}_{j + h}(\bbeta^*) - \mathbf{L}_{j + h}(\bbeta_0) \| = o_P(1),
\end{align}
then we have from \eqref{eq-m-est}
\begin{align*}
    \frac{1}{\sqrt{l_n}} \sum_{j = n - l_n + 1}^{n - h} Z_{j} (\tilde{Z}_{j + h} - Z_{j + h})
    &= \sqrt{\frac{l_n}{f_n}} \cdot \sqrt{f_n} (\hat{\bbeta} - \bbeta_0)^\T \cdot \frac{1}{l_n}\sum_{j = n - l_n + 1}^{n - h} Z_{j} \mathbf{L}_{j + h}(\bbeta_0) + o_P(1)\\
    &\gid \sqrt{k_{ra}} \mathbf{Q}^\T \E [Z_0 \mathbf{L}_h (\bbeta_0)],
\end{align*}
which follows via the ergodic theorem (for example, $Z_j \mathbf{L}_{j + h}$ is  a stationary sequence).
This establishes convergence of the second term in \eqref{eq-acf-re_exp}.
To prove \eqref{eq-fin-acf-z-l}, as $\|\hat{\bbeta} - \bbeta\| = o_P(1)$ it suffices to show
\begin{align*}
    \limsup_{\delta \downarrow 0} \limsup_{n \rightarrow \infty} \frac{1}{l_n} \sum_{j = n - l_n + 1}^{n - h} \E [| Z_j | \| \mathbf{L}_{j + h}(\bbeta^*) - \mathbf{L}_{j + h}(\bbeta_0) \| \ind\{\|\hat{\bbeta} - \bbeta_0\| \leq \delta\} ] = 0.
\end{align*}
We have for any $\delta > 0$ via Cauchy-Schwarz that
\begin{align*}
    &\limsup_{n \rightarrow \infty}\frac{1}{l_n} \sum_{j = n - l_n + 1}^{n - h} \E [| Z_j | \| \mathbf{L}_{j + h}(\bbeta^*) - \mathbf{L}_{j + h}(\bbeta_0) \| \ind\{\|\hat{\bbeta} - \bbeta_0\| \leq \delta\} ] \\
    & \;\;\;\; \leq \limsup_{n \rightarrow \infty} \frac{\sigma}{l_n} \sum_{j = n - l_n + 1}^{n - h} \Big( \E [ \| \mathbf{L}_{j + h}(\bbeta^*) - \mathbf{L}_{j + h}(\bbeta_0) \|^2 \ind\{\|\hat{\bbeta} - \bbeta_0\| \leq \delta\} ] \Big)^{1/2} \\
    & \;\;\;\; \leq \sigma \Big[ \E \sup_{\|\bbeta - \bbeta_0\| \leq \delta}  \| \mathbf{L}_{0}(\bbeta) - \mathbf{L}_{0}(\bbeta_0) \|^2 \Big]^{1/2},
\end{align*}
which goes to zero as $\delta \downarrow 0$ due to Assumption~\ref{cond-l}.
This completes the proof of \eqref{eq-fin-acf-z-l}.

Applying the same arguments for the third term in \eqref{eq-acf-re_exp} and noting $\E [Z_h \mathbf{L}_0(\bbeta_0)] = \mathbf{0}$ under causality as $h \geq 1$, we have
\begin{align*}
    \frac{1}{\sqrt{l_n}} \sum_{j = n - l_n + 1}^{n - h} (\hat{Z}_{j} - Z_{j}) Z_{j + h} = o_P(1).
\end{align*}
Similar arguments for the last term show that
\begin{align*}
    \frac{1}{\sqrt{l_n}} \sum_{j = n - l_n + 1}^{n - h} &(\hat{Z}_{j} - Z_{j}) (\hat{Z}_{j + h} - Z_{j + h}) \nonumber \\
    &= (\hat{\bbeta} - \bbeta_0)^\T \bigg( \frac{1}{l_n} \sum_{j = n - l_n + 1}^{n - h} \mathbf{L}_{j}(\bbeta^*_1) \mathbf{L}_{j + h}^\T(\bbeta^*_2) \bigg) \sqrt{l_n} (\hat{\bbeta} - \bbeta_0)
    + o_P(1),
\end{align*}
where $\bbeta^*_{\ell} = \bbeta + \varepsilon_{\ell}(\hat{\bbeta} - \bbeta_0)$, for some $\varepsilon_{\ell} \in [0, 1], \ell = 1,2$.
$\E [\mathbf{L}_0^\T(\bbeta_0) \mathbf{L}_h(\bbeta_0)]$ is finite by Cauchy-Schwarz and \eqref{eq-L-moments}.
The ergodic theorem and \eqref{eq-m-asymp-q} imply that above term and hence the fourth term in \eqref{eq-acf-re_exp} goes to zero in probability.
In summary, we obtain
\begin{align}
    \label{eq-acf-asymp-exp}
    \sqrt{l_n} \hat{\gamma}^{\hat{Z}}_{l_n}(h)
    = \sqrt{l_n} \hat{\gamma}^{Z}_{l_n}(h)
    + \E [Z_0 \mathbf{L}_h^\T(\bbeta_0)] \cdot \frac{\sqrt{l_n}}{f_n} \sum_{k = 1}^{f_n} \mathbf{m}(Z_{-\infty:k}; \bbeta_0)
    + o_P(1).
\end{align}
Next we consider joint convergence of
\begin{align}
    \label{eq-acf-joint-conv}
    \sqrt{l_n} \biggl( \hat{\gamma}_{l_n}^{Z} (h), \frac{1}{f_n} \sum_{k = 1}^{f_n} \mathbf{m}(Z_{-\infty:k}; \bbeta_0) \biggr).
\end{align}
Note that $\sqrt{l_n} \hat{\gamma}^{Z}_{l_n}(h) \gid \mathcal{N}(0, \sigma^4)$ by the central limit theorem and recall \eqref{eq-m-asymp-q} so that $\sqrt{l_n} (\hat{\bbeta} - \bbeta_0) \gid \sqrt{k_{ra}} \mathbf{Q}$.
The marginal convergence is guaranteed and we now consider the joint convergence.
Now if $k_{ov} = 0$, then the two terms are independent and we thus get joint convergence.
For the case $k_{ov} > 0$, the above independence argument can be applied for the last $n - f_n$ terms of $\hat{\gamma}^{Z}_{l_n}(h)$.
It thus suffices to show joint convergence of the overlapping terms, namely that
\begin{align}
    \label{eq-acf-joint-mgclt}
    \sqrt{l_n} \biggl(
        \frac{1}{l_n - h} \sum_{j = n - l_n + 1}^{f_n - h} Z_{j} Z_{j + h},
        \frac{1}{f_n} \sum_{k = 1}^{f_n} \mathbf{m}(Z_{-\infty:k}; \bbeta_0)
    \biggr)
\end{align}
is asymptotical normal.
Consider the triangular array with $n$th row given by
\begin{align*}
    \bzeta_{n j} := \frac{1}{\sqrt{n}} \Big( \ind\{j > n - l_n + h\} Z_{j - h} Z_{j}, \mathbf{m}(Z_{-\infty:j}; \bbeta_0) \Big) \text{ for } 1 \leq j \leq f_n.
\end{align*}
Let $\mathcal{F}_{n j}$ be the $\sigma$-algebra generated by $\{Z_k, k \leq j\}$.
It is easy to see that $\{\bzeta_{n j}\}_{1 \leq j \leq f_n}$ is adapted to $\{\mathcal{F}_{n j}\}_{1 \leq j \leq f_n}$ and is a martingale difference sequence as $\E[Z_{j - h} Z_{j} | \mathcal{F}_{j - 1}] = 0$ and $\E[\mathbf{m}(Z_{-\infty:j}; \bbeta_0) | \mathcal{F}_{j - 1}] = \mathbf{0}$.
Since $\E[Z_1^2] = \sigma^2 < \infty$ and $\E|\mathbf{m}(Z_{-\infty:0}; \bbeta_0)|^2 < \infty$, the martingale central limit theorem implies that $\sum_{j = 1}^{f_n} \bzeta_{n, j}$ converges to a Gaussian distribution; see Theorem 35.12 of \citep{Billingsley1995}.
This shows that the term in \eqref{eq-acf-joint-mgclt} converges jointly to a normal random vector.

With joint convergence justified, it remains to calculate the limiting cross-covariance between $\hat{\gamma}_{l_n}^{Z} (h)$ and $\frac{1}{f_n} \sum_{k = 1}^{f_n} \mathbf{m}(Z_{-\infty:k}; \bbeta_0)$, that is
\begin{align}
    \label{eq-fin-acf-cov-offdiag-lim}
    \lim_{n \rightarrow \infty} \frac{1}{f_n} \sum_{j = n - l_n + h + 1}^n \sum_{k = 1}^{f_n} \Cov\big( Z_{j - h} Z_j, \mathbf{m} (Z_{-\infty:k}; \bbeta_0)\big).
\end{align}
Now for any $j, k$ we have $\Cov\big( Z_{j - h} Z_j, \mathbf{m} (Z_{-\infty:k}; \bbeta_0)\big) = \E [ Z_{j - h} Z_j\mathbf{m}^\T(Z_{-\infty:k}; \bbeta_0) ]$.
Due to independence, for $j > k$ this expectation is zero.
Also for $j < k$, we have due to \eqref{eq-m-moments} that
\begin{align*}
    \E [ Z_{j - h} Z_j\mathbf{m}(Z_{-\infty:k}; \bbeta_0) ] = \E [ Z_{j - h} Z_j \E[ \mathbf{m}(Z_{-\infty:k}; \bbeta_0) | \mathcal{F}_{k-1} ] ] = \mathbf{0}.
\end{align*}
At $j = k$, the expectation equals $\E [ Z_0 Z_h \mathbf{m}(Z_{-\infty:h}; \bbeta_0) ]$.
Note that in the limit as $n \rightarrow \infty$, the term
\begin{align*}
    \frac{1}{f_n} \sum_{j = n - l_n + h + 1}^{n} \sum_{k = 1}^{f_n} \mathbb{I}\{j = k\}
\end{align*}
converges to the coefficient of overlap, $k_{ov}$.
Therefore the limit in \eqref{eq-fin-acf-cov-offdiag-lim} converges to $k_{ov} \bmu_{2 h}^\T$.
In total, from \eqref{eq-acf-asymp-exp} we conclude
\begin{align*}
    \sqrt{l_n} \hat{\gamma}_{l_n}^{\hat{Z}}(h) \gid \mathcal{N}(0, \sigma^4 \sigma_h^2),
\end{align*}
which completes the proof for the single lag case.

The above arguments can be easily extended for joint convergence at lags $1, \ldots, h$ (via the Cram\'{e}r-Wold device) and we have
\begin{align*}
    \sqrt{l_n} \Big( \hat{\gamma}^{\hat{Z}}_{l_n}(1), \ldots, \hat{\gamma}^{\hat{Z}}_{l_n}(h), \frac{1}{f_n} \sum_{k = 1}^{f_n} \mathbf{m}(Z_{-\infty:k}; \bbeta_0) \Big) \gid \mathcal{N} \bigg( \mathbf{0}, 
        \begin{pmatrix}
            \sigma^4 I_h & k_{ov} M_2 \\
            k_{ov} M_2^\T & k_{ra} \Var(\mathbf{Q})
        \end{pmatrix}
    \bigg).
\end{align*}
From this, \eqref{eq-fin-acf-var-cond} follows which completes the proof.

\subsection{Proof of Assumptions~\ref{cond-m-est}--\ref{cond-tilde-z} for ARMA Models}
\label{app-arma-assump}

Assumption~\ref{cond-m-est} is verified in Appendix~C of \citet{Davisetal2018}.
Furthermore, it was shown there that
\begin{align*}
    \mathbf{L}_0 (\bbeta) = \frac{1}{\theta(B)} \big( X_{-1}, \ldots, X_{-p}, Z_{-1}(\bbeta), \ldots, Z_{-q} (\bbeta) \big)
\end{align*}
Due to invertibility there exists a power series for $1/\theta(z)$ such that
\begin{align*}
    \frac{1}{\theta(z)} = \sum_{j = 0}^{\infty} \xi_j(\bbeta_0) z^j
\end{align*}
converges for all $|z| < 1 + \epsilon$, for some $\epsilon > 0$.
Then there exists a $\delta^* > 0$ such that for any $\|\bbeta - \bbeta_0\| \leq \delta^*$, $\sum_{j = 0}^{\infty} \xi_j(\bbeta) z^j$ converges for all $|z| < 1 + \epsilon/2$.
In particular, as $j \rightarrow \infty$ we have
\begin{align*}
    \xi_j(\bbeta)(1 + \epsilon/4)^j \rightarrow 0,
\end{align*}
and there exists $K > 0$ such that $|\pi_j(\bbeta)| \leq K r^j$, where $r = (1 + \epsilon/4)^{-1}$.
By similar arguments applied to the coefficients of the power series
\begin{align*}
    \frac{\phi(z)}{\theta(z)} = \sum_{j = 0}^{\infty} \pi_j(\bbeta_0) z^j,
\end{align*}
there exists $\delta^{**} > 0$, $\tilde{K} > 0$ and $0 < \tilde{r} < 1$ such that $|\pi_j(\bbeta)| \leq \tilde{K} \tilde{r}^j$ for all $j$ and for all $\|\bbeta - \bbeta\| \leq \delta^{**}$.
Then with $\delta_0 = \delta^* \wedge \delta^{**}$, for  $1 \leq \ell \leq p$ we have
\begin{align*}
    \E \sup_{\|\bbeta - \bbeta_0\| \leq \delta_0} \bigg( \frac{1}{\theta(B)} X_{-\ell} \bigg)^2
    &\leq \E \Big( \sum_{j = 0}^{\infty} K r^j |X_{-j - \ell}|  \Big)^2
    = K^2 \Big\| \sum_{j = 1}^{\infty} r^j X_{-j -\ell} \Big\|^2 \\
    &\leq K^2 \Big( \sum_{j = 1}^{\infty} r^j \| X_0 \| \Big)^2
    = K^2 \E X_0^2 / (1 - r)^2 < \infty,
\end{align*}
where $\|Y\| = \sqrt{\E Y^2}$ for a random variable $Y$.
Similarly for $1 \leq \ell \leq q$,
\begin{align*}
    \E \sup_{\|\bbeta - \bbeta_0\| \leq \delta_0} \bigg( \frac{1}{\theta(B)} Z_{-\ell}(\bbeta) \bigg)^2
    \leq \E \Big( \sum_{j, k = 0}^{\infty} K \tilde{K} r^j \tilde{r}^k |X_{-j - k - \ell}|  \Big)^2
    \leq \frac{K^2 \tilde{K}^2 \E X_0^2}{(1 - r)^2 (1 - \tilde{r})^2} < \infty.
\end{align*}
This completes the verification of Assumption~\ref{cond-l}.
For Assumption~\ref{cond-tilde-z}, we first claim
\begin{align}
    \label{eq-cond-tilde-etp}
    \sum_{j = 1}^{n} \sup_{\|\bbeta - \bbeta_0\| \leq \delta_0} | \hat{Z}_j(\bbeta) - Z_j(\bbeta) | = O_P(1).
\end{align}
This follows since for $n \geq 1$, we have
\begin{align*}
    \E  \sum_{j = 1}^{n} \sup_{\|\bbeta - \bbeta_0\| \leq \delta_0} | \hat{Z}_j(\bbeta) - Z_j(\bbeta) |
    &\leq \E  \sum_{j = 1}^{\infty} \sup_{\|\bbeta - \bbeta_0\| \leq \delta_0} | \hat{Z}_j(\bbeta) - Z_j(\bbeta) | \\
    &\leq K \E \sum_{j = 1}^{\infty} \sum_{t = j}^{\infty} \tilde{r}^t |X_{j - t}|
    = \tilde{K} \E |X_0| \tilde{r} / (1 - \tilde{r})^2 < \infty.
\end{align*}
As $\sqrt{n} \|\bbeta_n^\dagger - \bbeta_0\| = O_P(1)$, we have due to \eqref{eq-cond-tilde-etp} that
\begin{align*}
    \sum_{j = 1}^{n} | \hat{Z}_j(\bbeta_n^\dagger) - \tilde{Z}_j(\bbeta_n^\dagger) | = O_P(1).
\end{align*}
Therefore
\begin{align*}
    &\frac{1}{\sqrt{n}} \sum_{j = 1}^{n} |\hat{Z}_j (\bbeta_n^\dagger) - \tilde{Z}_j (\bbeta_n^\dagger)| = o_P(1),\\
    \frac{1}{\sqrt{n}} \sum_{j = 1}^{n} |\hat{Z}_j (\bbeta_n^\dagger) - \tilde{Z}_j (\bbeta_n^\dagger)|^2 &\leq \frac{1}{\sqrt{n}} \sum_{j = 1}^{n} |\hat{Z}_j (\bbeta_n^\dagger) - \tilde{Z}_j (\bbeta_n^\dagger)|\, \sum_{j = 1}^{n} |\hat{Z}_j (\bbeta_n^\dagger) - \tilde{Z}_j (\bbeta_n^\dagger)| = o_P(1).
\end{align*}

\subsection{Proof of Theorem~\ref{thm-fin-acf-sq-asymp}}
\label{app-acf-asymp-sq}

The asymptotics of the mean corrected autocovariance of the squared residuals are given by
\begin{align*}
    \hat{\gamma}_{l_n}^{\hat{Z}^2}(h) = \frac{1}{l_n} \sum_{j = n - l_n + 1}^{n - h} \hat{Z}_j^2 \hat{Z}_{j + h}^2 - \Big( \frac{1}{l_n} \sum_{j = n - l_n + 1}^{n} \hat{Z}_j^2 \Big)^2 + o_P(1 / \sqrt{l_n}) =: A_1 + A_2 + o_P(1/\sqrt{l_n}).
\end{align*}

For $A_1$ we have
\begin{align*}
    \sum_{j = n - l_n + 1}^{n - h} \hat{Z}_j^2 \hat{Z}_{j + h}^2
    &= \sum_{j = n - l_n + 1}^{n - h} (\hat{Z}_j^2 - Z_j^2) (\hat{Z}_{j + h}^2 - Z_{j + h}^2)
    + \sum_{j = n - l_n + 1}^{n - h} Z_j^2 (\hat{Z}_{j + h}^2 - Z_{j + h}^2) \\
    &+ \sum_{j = n - l_n + 1}^{n - h} (\hat{Z}_j^2 - Z_j^2) Z_{j + h}^2
    + \sum_{j = n - l_n + 1}^{n - h} Z_j^2 Z_{j + h}^2.
\end{align*}
Using Cauchy-Schwarz, Assumption~\ref{cond-tilde-z-2} and $\E[Z_0^4] < \infty$, we have for the third term
\begin{align*}
    \frac{1}{\sqrt{l_n}}  \sum_{j = n - l_n + 1}^{n - h} (\hat{Z}_j^2 - Z_j^2) Z_{j + h}^2 =
    \frac{1}{\sqrt{l_n}}  \sum_{j = n - l_n + 1}^{n - h} (\tilde{Z}_j^2 - Z_j^2) Z_{j + h}^2 + o_P(1).
\end{align*}
By Taylor's theorem
\begin{align*}
    \frac{1}{\sqrt{l_n}} \sum_{j = n - l_n + 1}^{n - h} (\tilde{Z}_j^2 - Z_j^2) Z_{j + h}^2 = 
    2 \sqrt{l_n} (\hat{\bbeta} - \bbeta_0)^\T \frac{1}{l_n} \sum_{j = n - l_n + 1}^{n - h} Z_j Z_{j + h}^2 \mathbf{L}_j(\bbeta^*),
\end{align*}
where $\bbeta^* = \bbeta_0 + \epsilon (\hat{\bbeta} - \bbeta_0)$ for some $\epsilon \in [0, 1]$.
Since $\E[Z_0^2 Z_h^4] < \infty$ we can show
\begin{align*}
    \frac{1}{l_n} \sum_{j = n - l_n + 1}^{n - h} | Z_j | Z_{j + h}^2 \| \mathbf{L}_{j}(\bbeta^*) - \mathbf{L}_{j}(\bbeta_0) \| = o_P(1),
\end{align*}
along the same lines as in the proof of \eqref{eq-fin-acf-z-l}.
This yields
\begin{align*}
    \frac{1}{\sqrt{l_n}} \sum_{j = n - l_n + 1}^{n - h} (\tilde{Z}_j^2 - Z_j^2) Z_{j + h}^2 = 
    2 \sqrt{l_n} (\hat{\bbeta} - \bbeta_0)^\T \frac{1}{l_n} \sum_{j = n - l_n + 1}^{n - h} Z_j Z_{j + h}^2 \mathbf{L}_j(\bbeta_0) + o_P(1).
\end{align*}
As $\{\mathbf{L}_j(\bbeta)\}$ is stationary and ergodic,
\begin{align*}
    \frac{1}{\sqrt{l_n}} \sum_{j = n - l_n + 1}^{n - h} (\tilde{Z}_j^2 - Z_j^2) Z_{j + h}^2
    = 2 \sqrt{l_n} (\hat{\bbeta} - \bbeta_0)^\T \E[Z_0 Z_h^2 \mathbf{L}_0(\bbeta_0)] + o_P(1),
\end{align*}
where we have used \eqref{eq-m-est}.
Thus
\begin{align*}
    \frac{1}{\sqrt{l_n}} \sum_{j = n - l_n + 1}^{n - h} (\hat{Z}_j^2 - Z_j^2) Z_{j + h}^2
    = 2 \sqrt{l_n} (\hat{\bbeta} - \bbeta_0)^\T \E[Z_0 Z_h^2 \mathbf{L}_0(\bbeta_0)] + o_P(1).
\end{align*}
Similarly,
\begin{align*}
    \frac{1}{\sqrt{l_n}} \sum_{j = n - l_n + 1}^{n - h} (\hat{Z}_{j + h}^2 - Z_{j + h}^2) Z_{j}^2
    = 2 \sqrt{l_n} (\hat{\bbeta} - \bbeta_0)^\T \E[Z_0^2 Z_h \mathbf{L}_h(\bbeta_0)] + o_P(1).
\end{align*}
We also have
\begin{align*}
    \frac{1}{\sqrt{l_n}} &\sum_{j = n - l_n + 1}^{n - h} (\tilde{Z}_j^2 - Z_j^2) (\tilde{Z}_{j + h}^2 - Z_{j + h}^2) \\
    &= 4 \sqrt{l_n} (\hat{\bbeta} - \bbeta_0) ^\T \E[ Z_0 Z_h \mathbf{L}_0(\bbeta_0) \mathbf{L}_h^\T(\bbeta_0) ] (\hat{\bbeta} - \bbeta_0) + o_P(1) = o_P(1).
\end{align*}
Similarly,
\begin{align*}
    \frac{1}{\sqrt{l_n}} \sum_{j = n - l_n + 1}^{n} (\hat{Z}_j^2 -  Z_j^2)
    &= \frac{1}{\sqrt{l_n}} \sum_{j = n - l_n + 1}^{n} (\tilde{Z_j^2} - Z_j^2)
    + \frac{1}{\sqrt{l_n}} \sum_{j = n - l_n + 1}^{n} (\hat{Z}_j^2 - \tilde{Z}_j^2) \\
    &= 2 \sqrt{l_n} (\hat{\bbeta} - \bbeta_0)^\T \E[Z_0 \mathbf{L}_0(\bbeta_0)] + o_P(1).
\end{align*}
We thus have
\begin{align*}
    \sqrt{l_n} &\bigg( \Big(\frac{1}{l_n} \sum_{j = n - l_n + 1}^{n} \hat{Z}_j^2 \Big)^2 - \sigma^4 \bigg) \\
    &= 2 \sigma^2 \sqrt{l_n} \bigg(\frac{1}{l_n} \sum_{j + n - l_n + 1}^n \hat{Z}_j^2 - \sigma^2 \bigg)+ o_P(1)\\
    &= 2 \sigma^2 \sqrt{l_n} \bigg(2(\hat{\bbeta} - \bbeta_0)^\T \E[Z_0 \mathbf{L}_0(\bbeta_0)] + \frac{1}{l_n} \sum_{j + n - l_n + 1}^n Z_j^2 - \sigma^2\bigg) + o_P(1).
\end{align*}
Then using $\E[Z_0 Z_h^2 \mathbf{L}_0(\bbeta_0)] = \sigma^2 \E[Z_0 \mathbf{L}_0(\bbeta_0)]$,
\begin{align*}
    \sqrt{l_n} \hat{\gamma}_{l_n}^{\hat{Z}^2} (h)
    &= \sqrt{l_n} \bigg( \frac{1}{l_n} \sum_{j = n - l_n + 1}^{n - h} Z_j^2 Z_{j + h}^2 - \sigma^4 \bigg)
    - 2 \sigma^2 \sqrt{l_n} \bigg( \frac{1}{l_n} \sum_{j = n - l_n + 1}^{n} Z_j^2 - \sigma^2 \bigg) \\
    &+ \sqrt{l_n} (\hat{\bbeta} - \bbeta_0)^\T \bmu_{1h}^{(2)} + o_P(1).
\end{align*}
To show convergence of $\sqrt{l_n} \hat{\gamma}_{l_n}^{\hat{Z}^2} (h)$, it thus suffices to show convergence of
\begin{align*}
    \sqrt{l_n} \bigg( \frac{1}{l_n} \sum_{j = n - l_n + 1}^{n - h} Z_j^2 Z_{j + h}^2 - \sigma^4, \frac{1}{l_n} \sum_{j = n - l_n + 1}^{n} Z_j^2 - \sigma^2, \frac{1}{f_n} \sum_{k = 1}^{f_n} \mathbf{m}(Z_{-\infty:k}; \bbeta_0) \bigg).
\end{align*}
Marginally, it is easy to see that
\begin{align*}
    \frac{1}{\sqrt{l_n}} \sum_{j = n - l_n + 1}^{n - h} (Z_j^2 Z_{j + h}^2 - \sigma^4) &\gid \mathcal{N}(0, \E^2[Z_0^4] + 2\sigma^4 \E[Z_0^4] - 3\sigma^8), \\
    \frac{1}{\sqrt{l_n}} \sum_{j = n - l_n + 1}^{n} (Z_j^2 - \sigma^2) &\gid \mathcal{N}(0, \E[Z_0^4] - \sigma^4), \\
    \sqrt{l_n} \frac{1}{f_n} \sum_{k = 1}^{f_n} \mathbf{m}(Z_{-\infty:k}; \bbeta_0) &\gid \mathcal{N}(\mathbf{0}, k_{ra} \Var(\mathbf{Q})).
\end{align*}
The joint convergence to a Gaussian limit can be obtained by applying the martingale central limit theorem, as is done in the proof of Theorem~\ref{thm-fin-acf-asymp}.
It is easy to see that the cross-covariance between the first two terms converges to $2 \sigma^2 (\E[Z_0^4] - \sigma^4)$.
Consider now the cross-covariance between the first and third terms,
\begin{align*}
    \lim_{n \rightarrow \infty} \frac{1}{f_n} \sum_{j = n - l_n + 1}^{n - h} \sum_{k = 1}^{f_n} \Cov\big( Z_{j}^2 Z_{j + h}^2, \mathbf{m} (Z_{-\infty:k}; \bbeta_0)\big).
\end{align*}
For any $j > k$, the two terms are independent and the covariance is zero.
If $j < k$ and $j + h \neq k$, the covariance is zero yet again as $\E[\mathbf{m}(Z_{-\infty:k}; \bbeta_0) | \mathcal{F}_{k - 1})] = \mathbf{0}$.
Thus the covariance equals
\begin{align*}
    \ind\{j = k\} \sigma^2 \E[Z_0^2 \mathbf{m}^\T(Z_{-\infty:0}; \bbeta_0)] + \ind\{j + h = k\} \E[Z_0^2 Z_h^2 \mathbf{m}^\T(Z_{-\infty:h}; \bbeta_0)].
\end{align*}
Taking the limit we get $k_{ov} \E[\sigma^2 Z_0^2 \mathbf{m}^\T(Z_{-\infty:0}; \bbeta_0) + Z_0^2 Z_h^2 \mathbf{m}^\T(Z_{-\infty:h}; \bbeta_0)]$.
Similarly, the limiting cross-covariance between the second and third terms can be shown to be $k_{ov} \E[Z_0^2 \mathbf{m}^\T(Z_{-\infty:0}; \bbeta_0)]$.
This completes the proof.

\subsection{Proof of Assumptions~\ref{cond-m-est}--\ref{cond-tilde-z-2} for GARCH Models}
\label{app-garch-assump}
Assumption~\ref{cond-m-est} follows from Theorem 4.2 of \citet{Berkesetal2003b}.
For Assumption~\ref{cond-l}, Lemma~3.1 of \citet{KulpergerYu2005} yields
\begin{align*}
    \E \bigg( \sup_{\mathbf{u} \in \mathbf{\Theta}} \bigg\| \frac{\partial \log \sigma_j^2(\mathbf{u})}{\partial \mathbf{u}} \bigg\| \bigg  )^k < \infty,
\end{align*}
for any $k > 0$.
Applying Cauchy-Schwarz we get
\begin{align*}
    \E \sup_{\mathbf{u} \in \mathbf{\Theta}} \| \mathbf{L}_0 (\mathbf{u}) \|^2
    = \E \bigg[ \frac{1}{4} Z_0^2 \sup_{\mathbf{u} \in \mathbf{\Theta}} \bigg\| \frac{\partial \log \sigma_j^2(\mathbf{u})}{\partial \mathbf{u}}  \bigg\|^2 \bigg]
    \leq \frac{1}{4} \bigg( \E [Z_0^4] \E \bigg[ \sup_{\mathbf{u} \in \mathbf{\Theta}} \bigg\|  \frac{\partial \log \sigma_j^2(\mathbf{u})}{\partial \mathbf{u}}  \bigg\| \bigg]^4 \bigg)^{1/2}
    < \infty,
\end{align*}
as required.
It remains to show Assumptions~\ref{cond-tilde-z}--\ref{cond-tilde-z-2}.
We have
\begin{align*}
    \hat{Z}_j (\bbeta_n^\dagger) = \tilde{Z}_j (\bbeta_n^\dagger) \left( 1 + \frac{\tilde{\sigma}_j (\bbeta_n^\dagger) - \hat{\sigma}_j (\bbeta_n^\dagger)}{\hat{\sigma}_j(\bbeta_n^\dagger)} \right).
\end{align*}
With $i = k = 1, 2$ in Equation~(3.2) of \citet{KulpergerYu2005}, it follows that
\begin{align*}
    \sum_{j = 1}^{n} |\hat{Z}_j^k (\bbeta_n^\dagger) - \tilde{Z}_j^k (\bbeta_n^\dagger)| = O_P(1).
\end{align*}
Assumptions~\ref{cond-tilde-z}--\ref{cond-tilde-z-2} now follow since
\begin{align*}
    &\frac{1}{\sqrt{n}} \sum_{j = 1}^{n} |\hat{Z}_j^k (\bbeta_n^\dagger) - \tilde{Z}_j^k (\bbeta_n^\dagger)| = o_P(1),\\
    \frac{1}{\sqrt{n}} \sum_{j = 1}^{n} |\hat{Z}_j^k (\bbeta_n^\dagger) - \tilde{Z}_j^k (\bbeta_n^\dagger)|^2 &\leq \frac{1}{\sqrt{n}} \sum_{j = 1}^{n} |\hat{Z}_j^k (\bbeta_n^\dagger) - \tilde{Z}_j^k (\bbeta_n^\dagger)|\, \sum_{j = 1}^{n} |\hat{Z}_j^k (\bbeta_n^\dagger) - \tilde{Z}_j^k (\bbeta_n^\dagger)| = o_P(1),
\end{align*}
for $k = 1,2$.

\subsection{Proof of Theorem~\ref{thm-fin-dcf-asymp}}
\label{app-dcf-asymp}

Write
\begin{align*}
    l_n \hat{T}_{l_n}^{\hat{Z}}(h) =: \| \sqrt{l_n}C_{l_n h}^{\hat Z} \|^2_\mu = \| \sqrt{l_n} C_{l_n h}^{\hat Z} - \sqrt{l_n} C_{l_n h}^{Z} + \sqrt{l_n} C_{l_n h}^{Z} \|^2_\mu,
\end{align*}
where
\begin{align*}
    C_{l_n h}^{\hat{Z}}(s, t) &= \frac{1}{l_n} \sum_{j = n - l_n + 1}^{n - h} e^{i s \hat{Z}_{j} + i t \hat{Z}_{j + h}} - \frac{1}{l_n} \sum_{j = n - l_n + 1}^{n - h} e^{i s \hat{Z}_{j}} \frac{1}{l_n} \sum_{j = n - l_n + 1}^{n - h} e^{i t \hat{Z}_{j + h}},\\
    C_{l_n h}^{Z}(s, t) &= \frac{1}{l_n} \sum_{j = n - l_n + 1}^{n - h} e^{i s Z_{j} + i t Z_{j + h}} - \frac{1}{l_n} \sum_{j = n - l_n + 1}^{n - h} e^{i s Z_{j}} \frac{1}{l_n} \sum_{j = n - l_n + 1}^{n - h} e^{i t Z_{j + h}}.
\end{align*}
We will first show that
\begin{align}
    \label{eq-fin-adcf-chfn}
    \sqrt{l_n} C_{l_n h}^{\hat Z} \gid G_h + \xi_h,
\end{align}
where the convergence is in $\mathcal{C}(K)$, $K \subset \R^2$ is a compact set and $G_h + \xi_h$ has covariance function given by \eqref{eq-fin-dcf-cov}.
%
%
Let
\begin{align*}
    E_{l_n}(s, t) =  \frac{1}{\sqrt{l_n}} \sum_{j = n - l_n + 1}^{n - h} \big( e^{i s \hat{Z}_j + i t \hat{Z}_{j + h}} - e^{i s Z_j + i t Z_{j + h}} \big),
\end{align*}
so that
\begin{align*}
    \sqrt{l_n} ( C_{l_n h}^{\hat{Z}}(s, t) &- C_{l_n h}^{Z}(s, t) ) \nonumber \\
    &= \frac{1}{\sqrt{l_n}} \sum_{j = n - l_n + 1}^{n - h} \big( e^{i s \hat{Z}_j + i t \hat{Z}_{j + h}} - e^{i s Z_j + i t Z_{j + h}} \big) \nonumber \\
    &- \frac{1}{\sqrt{l_n}} \sum_{j = n - l_n + 1}^{n - h} \big(e^{i s \hat{Z}_j} - e^{i s Z_j} \big) \frac{1}{l_n} \sum_{j = n - l_n + 1}^{n - h} e^{i t Z_{j + h}} \nonumber \\
    &- \frac{1}{l_n} \sum_{j = n - l_n + 1}^{n - h} e^{i s \hat{Z}_j} \frac{1}{\sqrt{l_n}} \sum_{j =n - l_n + 1}^{n - h} \big(e^{i t \hat{Z}_{j + h}} - e^{i t Z_{j + h}} \big)\nonumber \\
    &= E_{l_n}(s, t) - E_{l_n}(s, 0) \frac{1}{l_n}\sum_{j = n - l_n + 1}^{n - h} e^{i t Z_{j + h}} - E_{l_n}(0, t) \frac{1}{l_n} \sum_{j = n - l_n + 1}^{n - h} e^{i s \hat{Z}_j}.
\end{align*}
We consider first the asymptotics of $E_{l_n}(s, t)$.
Using same arguments as in the proof of Proposition A.1 in \citet{DavisWan2020}, write
\begin{align*}
    E_{l_n}(s, t)
    &= \frac{1}{l_n} \sum_{j = n - l_n + 1}^{n - h} e^{i s Z_j + i t Z_{j + h}} \big(i s \sqrt{l_n}(\hat{Z}_j -  \tilde{Z}_j)+it\sqrt{l_n}(\hat{Z}_{j+h}- \tilde{Z}_{j+h})\big) \nonumber \\
    &+ \frac{1}{l_n} \sum_{j = n - l_n + 1}^{n - h} e^{i s Z_j + i t Z_{j + h}} \big(i s \sqrt{l_n}(\tilde{Z}_j - Z_j) + i t \sqrt{l_n} (\tilde{Z}_{j + h} - Z_{j + h})\big) + o_P(1) \nonumber \\
    &=: E_{l_n1} (s, t) + E_{l_n2} (s, t) + o_P(1)
\end{align*}
Due to Assumption~\ref{cond-tilde-z},
\begin{align*}
    |E_{l_n1}(s, t)| \leq |s| \frac{1}{\sqrt{l_n}} \sum_{j = n - l_n + 1}^{n-h} |\hat{Z}_j -  \tilde{Z}_j| + |t| \frac{1}{\sqrt{l_n}} \sum_{j = n - l_n + 1}^{n - h} |\hat{Z}_{j + h} -  \tilde{Z}_{j + h}| \gip 0, \;\; \text{ in } \mathcal{C}(K).
\end{align*}
Applying Taylor's theorem about $\bbeta_0$ we get
\begin{align*}
    E_{l_n2}(s, t)
    = \sqrt{l_n} (\hat{\bbeta} - \bbeta_0)^\T \frac{1}{l_n} \sum_{j = n - l_n + 1}^{n - h} e^{i s Z_j + i t Z_{j + h}} \big(i s \mathbf{L}_j (\bbeta^*) + i t \mathbf{L}_{j + h}(\bbeta^*) \big),
\end{align*}
where $\bbeta^* = \bbeta_0 + \varepsilon (\hat{\bbeta} - \bbeta_0)$ for some $0 \leq \varepsilon \leq 1$.
From the proof of \eqref{eq-fin-acf-z-l} we have
\begin{align*}
    \frac{1}{l_n} \sum_{j = n - l_n + 1}^{n - h} \| \mathbf{L}_j (\bbeta^*) - \mathbf{L}_j (\bbeta_0) \| = o_P(1),
\end{align*}
so that on $\mathcal{C}(K)$
\begin{align*}
    E_{l_n2}(s, t)
    = \sqrt{l_n} (\hat{\bbeta} - \bbeta_0)^\T \frac{1}{l_n} \sum_{j = n - l_n + 1}^{n - h} e^{i s Z_j + i t Z_{j + h}} \big(i s \mathbf{L}_j (\bbeta_0) + i t \mathbf{L}_{j + h}(\bbeta_0) \big) + o_P(1).
\end{align*}
Since $\mathbf{L}_j(\bbeta)$ is stationary and ergodic, in view of the uniform ergodic theorem,
\begin{align*}
    \frac{1}{l_n} \sum_{j = n - l_n + 1}^{n-h} e^{i s Z_j + i t Z_{j + h}} &\big( i s \mathbf{L}_j (\bbeta^*) + i t \mathbf{L}_{j + h}(\bbeta^*) \big) \nonumber \\
    &\gip \E\big[ e^{i s Z_j + i t Z_{j + h}} \big( i s \mathbf{L}_j (\bbeta_0) + i t \mathbf{L}_{j + h}(\bbeta_0) \big) \big] =: \mathbf{E}_h(s,t), \;\; \text{ in } \mathcal{C}(K).
\end{align*}
Hence 
\begin{align*}
    E_{l_n}(s, t) = \sqrt{l_n} (\hat{\bbeta} - \bbeta_0)^\T  \mathbf{E}_h(s, t) + o_P(1), \;\; \text{ in } \mathcal{C}(K).
\end{align*}
Note that
\begin{align*}
    \frac{1}{l_n} \sum_{j = n - l_n + 1}^{n - h} e^{i t Z_{j + h}} \gip \varphi_Z(t), \;\; \text{ in } \mathcal{C}(K),
\end{align*}
and
\begin{align*}
    \frac{1}{l_n} \sum_{j = n - l_n + 1}^{n - h} e^{i s \hat{Z}_j}
    = \frac{1}{l_n} \sum_{j = n - l_n + 1}^{n - h} e^{i s Z_j} + \frac{1}{\sqrt{l_n}} E_{l_n}(s, 0)  \gip \varphi_Z(s), \;\; \text{ in } \mathcal{C}(K).
\end{align*}
Equation A.3 of \citet{DavisWan2020} gives $\mathbf{E}_h(s,t) - \mathbf{E}_h(s,0) \varphi_Z(t) - \mathbf{E}_h(0,t) \varphi_Z(s) = \bnu_{1 h}(s, t)$.
We thus have
\begin{align*}
    \sqrt{l_n} (C_{l_n h}^{\hat{Z}} - C_{l_n h}^Z) = \sqrt{l_n} (\hat{\bbeta} - \bbeta_0)^\T \bnu_{1 h}(s, t) + o_P(1), \;\; \text{ in } \mathcal{C}(K).
\end{align*}
It is shown in Theorem 1 of \citet{Davisetal2018} that
\begin{align}
    \label{eq-fin-dcf-C_sn_asymp}
    \sqrt{l_n} C_{l_n h}^{Z} = \sqrt{l_n} \tilde{C}_{l_n h}^{Z} + o_P(1) \gid G_h \;\; \text{ in } \mathcal{C}(K),
\end{align}
where
\begin{align*}
    \tilde{C}_{l_n h}^{Z} = \frac{1}{l_n} \sum_{j = n - l_n + 1}^{n - h} (e^{isZ_j} - \varphi_Z(s)) (e^{itZ_{j + h}} - \varphi_Z(t)).
\end{align*}
In total we have
\begin{align}
    \sqrt{l_n} C_{l_n h}^{\hat{Z}} = \sqrt{l_n} (\hat{\bbeta} - \bbeta_0)^\T \bnu_{1 h}(s, t) + \sqrt{l_n} \tilde{C}_{l_n h}^{Z} + o_P(1), \;\; \text{ in } \mathcal{C}(K).
\end{align}
To show convergence to a Gaussian, it suffices to show joint convergence of $\sqrt{l_n}(\hat{\bbeta} - \bbeta_0, \tilde{C}_{l_n h})$.
We now establish joint convergence of
\begin{align}
    \label{eq-fin-dcf-mgclt}
    \sqrt{l_n} \biggl( \frac{1}{f_n} \sum_{k = 1}^{f_n} \mathbf{m}(Z_{-\infty:k}; \bbeta_0), \frac{1}{l_n} \sum_{j = n - l_n + 1}^{f_n - h} (e^{isZ_j} - \varphi_Z(s)) (e^{itZ_{j + h} - \varphi_Z(t)}) \biggr).
\end{align}
Consider the triangular array with $n$th row given by
\begin{align*}
    \bvarrho_{n j} := \frac{1}{\sqrt{n}} \Big(\mathbf{m}(Z_{-\infty:j}; \bbeta_0),  \ind\{j > n - l_n\} (e^{isZ_{j - h}} - \varphi_Z(s)) (e^{itZ_{j}} - \varphi_Z(t)) \Big) \;\; \text{ for } 1 \leq j \leq f_n.
\end{align*}
Let $\mathcal{F}_{n j}$ be the $\sigma$-algebra generated by $\{Z_k, k \leq j\}$.
It is easy to see that $\{\bvarrho_{n j}\}_{1 \leq j \leq f_n}$ is adapted to $\{\mathcal{F}_{n j}\}_{1 \leq j \leq f_n}$ and is a martingale difference sequence as
\begin{align*}
    \E[(e^{isZ_{j - h}} - \varphi_Z(s)) (e^{itZ_{j}} - \varphi_Z(t)) | \mathcal{F}_{j - 1}] = 0
\end{align*}
and $\E[\mathbf{m}(Z_{-\infty:j}; \bbeta_0) | \mathcal{F}_{j - 1}] = \mathbf{0}$.
Due to the martingale central limit theorem, we have $\sum_{j = 1}^{f_n} \bvarrho_{n j}$ converging to a Gaussian distribution; see Theorem 35.12 of \citet{Billingsley1995}.
This shows joint convergence of \eqref{eq-fin-dcf-mgclt}.

To justify \eqref{eq-fin-dcf-cov} it suffices to show $\Cov \bigl( \sqrt{k_{ra}} \mathbf{Q}, G_h(s, t) \bigr) = k_{ov} \overline{\bnu_{2 h}(s, t)}$.
To this end, we evaluate the following limit
\begin{align*}
    \lim_{n\rightarrow\infty} \frac{1}{l_n} \sum_{j = n - l_n + h + 1}^{n} \sum_{k = 1}^{f_n} \Cov \bigl( \mathbf{m}(Z_{-\infty:k}; \bbeta_0), \big(e^{i s Z_{j - h}} - \varphi_Z(s)\big) \big(e^{i t Z_{j}} - \varphi_Z(t)\big) \bigr).
\end{align*}
Now for any $j, k$ we have $ \Cov \bigl( \mathbf{m}(Z_{-\infty:k}; \bbeta_0), \big(e^{i s Z_{j - h}} - \varphi_Z(s)\big) \big(e^{i t Z_{j}} - \varphi_Z(t)\big) \bigr)$.
Due to independence, for $j > k$ this covariance is zero.
Also for $j < k$, we have due to \eqref{eq-m-moments} that
\begin{align*}
    \E \big[ \mathbf{m}(Z_{-\infty:k}; \bbeta_0) &\big(\overline{e^{i s Z_{j - h}} - \varphi_Z(s)}\big) \big(\overline{e^{i t Z_{j}} - \varphi_Z(t)}\big) \big] \\
    &= \E \big[ \big(\overline{e^{i s Z_{j - h}} - \varphi_Z(s)}\big) \big(\overline{e^{i t Z_{j}} - \varphi_Z(t)}\big) \E \big[ \mathbf{m}(Z_{-\infty:k}; \bbeta_0) \big| \mathcal{F}_{k - 1} \big] \big]
    = \mathbf{0}.
\end{align*}
At $j = k$, the covariance equals $\overline{\bnu_{2 h}(s, t)}$ and hence the limit evaluates to $k_{ov} \overline{\bnu_{2 h}(s, t)}$.

The continuous mapping theorem then implies that
\begin{align*}
    l_n \int_{K_{\delta}} |C_{l_n h}^{\hat{Z}}|^2 d\mu \gid \int_{K_{\delta}} |G_h + \xi_h|^2 d\mu
\end{align*}
where
\begin{align*}
    K_{\delta} = \{ (s, t) \in \R^2 : \delta \leq |s| \leq 1/\delta, \; \delta \leq |t| \leq 1/\delta \},\;\;\delta \in (0, 1).
\end{align*}
To complete the proof it suffices to show for any $\varepsilon > 0$ that
\begin{align*}
    \lim_{\delta \rightarrow 0} \limsup_{n \rightarrow \infty} \mathbb{P} \bigg( \int_{K_{\delta}^c} \big| \sqrt{l_n} C_{l_n h}^{\hat{Z}} \big|^2 d\mu > \varepsilon \bigg) = 0 \\
    \lim_{\delta \rightarrow 0} \mathbb{P} \bigg(\int_{K_{\delta}^c} |G_h + \xi_h|^2 d\mu(s, t) > \varepsilon \bigg) = 0.
\end{align*}
The latter follows from Proposition A.3 of \citet{DavisWan2020}.

Following the notation of Proposition A.2 of \citet{DavisWan2020} we have
\begin{align*}
    C_{l_n h}^{\hat Z} - C_{l_n h}^{Z}
    &= \frac{1}{l_n} \sum_{j = n - l_n + 1}^{n - h} A_j B_j
    - \frac{1}{l_n} \sum_{j = n - l_n + 1}^{n - h} A_j \frac{1}{l_n} \sum_{j = n - l_n + 1}^{n - h} B_j \nonumber \\
    &- \frac{1}{l_n} \sum_{j = n - l_n + 1}^{n - h} U_j \frac{1}{l_n} \sum_{j = n - l_n + 1}^{n - h} B_j
    - \frac{1}{l_n} \sum_{j = n - l_n + 1}^{n - h} V_j\frac{1}{l_n} \sum_{j = n - l_n + 1}^{n - h} A_j \nonumber \\
    &+  \frac{1}{n} \sum_{j = n - l_n + 1}^{n - h} U_j B_j  + \frac{1}{l_n} \sum_{j = n - l_n + 1}^{n - h} V_jA_j \nonumber \\
    &=: \sum_{m = 1}^6 I_{l_n m}(s,t),
\end{align*}
where
\begin{align*}
    U_j = e^{isZ_j} - \varphi_Z(s),\quad V_j =  e^{itZ_{j + h}} - \varphi_Z(t), \quad A_j = e^{is\hat{Z}_j} - e^{isZ_j}, \quad B_j = e^{it\hat{Z}_{j + h}} - e^{itZ_{j + h}}.
\end{align*}
The same arguments show for $a = 1,2$ that
\begin{align*}
    \frac{1}{\sqrt{l_n}} \sum_{j = n - l_n + 1}^{n - h} |\hat{Z}_j - Z_j|^{a} 
    &\leq c \Big( \frac{1}{\sqrt{l_n}}\sum_{j = n - l_n + 1}^{n - h} |\hat{Z}_j - \tilde Z_j|^{a} + \frac{1}{\sqrt{l_n}} \sum_{j = n - l_n + 1}^{n - h} |\tilde{Z}_j-Z_j|^{a} \Big) \nonumber \\
    &\leq o_P(1) + c \sqrt{\frac{l_n}{f_n}} \frac{1}{f_n^{(a - 1) / 2}} \|\sqrt{f_n}(\hat{\bbeta} - \bbeta)\|^{a} \frac{1}{l_n} \sum_{j = n - l_n + 1}^{n}  \|\mathbf{L}_j(\bbeta^*)\|^{a} \nonumber \\
    &= O_P(1).
\end{align*}
Thus
\begin{align*}
    l_n |I_{l_n 1}(s, t)|^2 \leq \big( (1 \wedge |s|^2) (1 \wedge |t|^2) + (s^2 + t^2) \mathbb{I}(|s| \wedge |t| > 1) \big) O_P(1),
\end{align*}
where the $O_P(1)$ is free of $(s, t)$.
Therefore
\begin{align*}
    \lim_{\delta \rightarrow 0} \limsup_{n \rightarrow \infty} \mathbb{P} \bigg( \int_{K_\delta^c} l_n |I_{l_n 1}(s, t)|^2 d\mu(s, t) > \varepsilon \bigg) = 0.
\end{align*}
Similar arguments show that $l_n |I_{l_n 2}(s, t)|^2$ is bounded by $\min(|s|^2,|t|^2,|s t|^2) O_P(1)$, $l_n |I_{l_n 3}(s, t)|^2$ and $l_n |I_{l_n 5}(s, t)|^2$ are bounded by $\min(|t|^2,|s t|^2) O_P(1)$, and $l_n |I_{l_n 4}(s,t)|^2$ and $l_n |I_{l_n 6}(s,t)|^2$ are bounded by $\min(|s|^2,|st|^2) O_P(1)$.
This completes the proof. 


\subsection{Proof of Lemma~\ref{thm-stoch}}
\label{app-chfn}


Assume that $\tau_Z(t) \leq 1$ for all $t$ and $k_{ra} \geq 2 k_{ov}$.
Consider the compact set $K_{\delta} = \{(s,t) \in \R^2 : 1/\delta \leq |s|, |t| \leq \delta\}$ and fix a partition $\{(s_j,t_j)\}_{j=1}^m$ of $K_{\delta}$.
Consider the Gaussian random vectors $\mathbf{X}, \mathbf{Y} \in \R^m$ with components $X_j = (G_h + \xi_h) (s_j, t_j)$ and $Y_j = G_h (s_j, t_j)$.
Due to Example 1.A.28 in \citet{ShakedShanthikumar2007}, to show $\|\mathbf{X}\|^2 \succeq \|\mathbf{Y}\|^2$ it suffices to prove $\Cov(\mathbf{X}, \mathbf{X}) - \Cov(\mathbf{Y}, \mathbf{Y})$ is nonnegative definite.
To this end for $\mathbf{a} \in \R^m$ we have from \eqref{eq-fin-dcf-cov-cond} we have
\begin{align*}
  \mathbf{a}^\T (\Cov(\mathbf{X}) -  \Cov(\mathbf{Y})) \mathbf{a} 
  &\geq (k_{ra} - 2k_{ov}) \sum_{i, j = 1}^{m} a_i a_j \bnu_{1 h}^\T(s_i, t_i) \Var(\mathbf{Q}) \overline{\bnu_{1 h}(s_j, t_j)} \\
  &= (k_{ra} - 2k_{ov}) \bigg(\sum_{i = 1}^{m} a_i \bnu_{1 h}^\T(s_i, t_i) \bigg) \Var(\mathbf{Q}) \overline{\bigg(\sum_{i = 1}^{m} a_i \bnu_{1 h}(s_i, t_i) \bigg)}.
\end{align*}
The nonnegative definiteness of $\Cov(\mathbf{X}, \mathbf{X}) - \Cov(\mathbf{Y}, \mathbf{Y})$ follows from nonnegative definiteness of $\Var(\mathbf{Q})$.
Thus $\|\mathbf{X}\|^2 \succeq \|\mathbf{Y}\|^2$; this further implies $\|A \mathbf{X}\|^2 \succeq \|A \mathbf{Y}\|^2$ for any matrix $A \in \R^{m \times m}$, since $\Cov(A \mathbf{X}) - \Cov(A \mathbf{Y}) = A (\Cov(\mathbf{X}) - \Cov(\mathbf{Y}))A^\T$.
In particular taking $A$ to be diagonal matrix with entries $A_{jj} = \sqrt{f_{\mu}(s_j, t_j)}$, where $f_{\mu}$ is the density function corresponding to $\mu$, we have
\begin{align*}
  \sum_{j=1}^m \|(G_h + \xi_h) (s_j, t_j)\|^2 f_{\mu} (s_j, t_j) \succeq  \sum_{j = 1}^m \|G_h (s_j, t_j)\|^2 f_{\mu} (s_j, t_j).
\end{align*}
Taking finer and finer partitions, since $\{(s_j,t_j)\}_{j=1}^m$ was arbitrary, the Riemann sums converge due to \eqref{eq-weight} and we have
\begin{align*}
  \int_{K_{\delta}} \|G_h + \xi_h\|^2 d\mu \succeq \int_{K_{\delta}} \|G_h\|^2 d\mu.
\end{align*}
Taking $\delta \downarrow 0$ completes the proof.
The proof for the case $\tau_Z(t) \geq 1$ for all $t$ and $k_{ra} \leq 2k_{ov}$ follows by the same arguments with $X_j = G_h (s_j, t_j)$ and $Y_j = (G_h + \xi_h) (s_j, t_j)$.

\subsection{Proof of Lemma~\ref{thm-chfn-ineq}}
\label{app-chfn-ineq}

Note that $t \mapsto \varphi_Z'(t)/(t \varphi_Z(t))$ is an even function which approaches $-\sigma^2$ as $t$ goes to 0.
Thus it suffices to consider $t > 0$.
The Laplace distribution with variance $\sigma^2$ has characteristic function $\varphi_Z(t) = (1+\sigma^2t^2/2)^{-1}$.
A simple calculation shows
\begin{align*}
    \varphi_Z'(t)/(t \varphi_Z(t)) = -\sigma^2 \varphi_Z(t) \geq -\sigma^2.
\end{align*}
Consider now the case where $Z$ has a student's t distribution.
Due to Equation (1.2) in \citet{Gaunt2021}
\begin{align*}
    \varphi_Z(t) = \frac{K_{\nu / 2}(\sqrt{\nu}|t|) (\sqrt{\nu}|t|)^{\nu / 2}}{\Gamma({\nu}/{2}) 2^{\nu / 2 - 1}}, \;\;\;\; t \in \mathbb{R},
\end{align*}
where $K_{\nu / 2}$ is a modified Bessel function of the second kind.
Now $\frac{d}{dx} (x^{\nu} K_{\nu}(x)) = - x^{\nu} K_{\nu - 1}(x)$ 
for all $x$ so that
\begin{align*}
    \frac{\varphi_Z'(t)}{\sigma^2 t \varphi_Z(t)} = (\nu - 2) \cdot \frac{-K_{\nu / 2 - 1}(\sqrt{\nu} |t|)}{\sqrt{\nu} |t| K_{\nu / 2}(\sqrt{\nu} |t|)}.
\end{align*}
For $t > 0$, we have 
\begin{align*}
    (\nu-2) &\cdot \frac{d}{dt} \bigg( \frac{-(\sqrt{\nu} t)^{\nu / 2 - 1} K_{\nu / 2 - 1}(\sqrt{\nu} t)}{(\sqrt{\nu} t)^{\nu / 2} K_{\nu / 2}(\sqrt{\nu} t)} \bigg) \\
    &= \frac{\nu - 2}{(\sqrt{\nu} t)^{\nu} K_{\nu / 2}^2(\sqrt{\nu}t)} \cdot \big((\sqrt{\nu} t)^{\nu - 1} K_{\nu / 2 - 2}(\sqrt{\nu} t)  K_{\nu / 2}(\sqrt{\nu} t) - (\sqrt{\nu} t)^{\nu - 1} K_{\nu / 2 - 1}^2(\sqrt{\nu} t)\big) \\
    &= \frac{\nu - 2}{\sqrt{\nu} t K_{\nu / 2}^2(\sqrt{\nu}t)} \cdot \big(K_{\nu / 2 - 2}(\sqrt{\nu} t) K_{\nu / 2}(\sqrt{\nu} t) - K_{\nu/2 - 1}^2(\sqrt{\nu} t)\big)
    > 0,
\end{align*}
where the last inequality follows from the Turan type inequality for modified Bessel functions that states $K_{\nu / 2 - 2}(x) K_{\nu / 2}(x) - K_{\nu/2 - 1}^2(x) > 0$ for all $x \in \R$; see \citet{BariczPonnusamy2013} for details.
As stated previously, the symmetry around zero establishes the result for $t < 0$ and hence the proof is complete.

\bibliographystyle{plainnat}
\bibliography{sspl.bib}

\begin{thebibliography}{18}
\providecommand{\natexlab}[1]{#1}
\providecommand{\url}[1]{\texttt{#1}}
\expandafter\ifx\csname urlstyle\endcsname\relax
  \providecommand{\doi}[1]{doi: #1}\else
  \providecommand{\doi}{doi: \begingroup \urlstyle{rm}\Url}\fi

\bibitem[Baricz and Ponnusamy(2013)]{BariczPonnusamy2013}
\'{A}rp\'{a}d Baricz and Saminathan Ponnusamy.
\newblock On {T}ur\'{a}n type inequalities for modified {B}essel functions.
\newblock \emph{Proc. Amer. Math. Soc.}, 141\penalty0 (2):\penalty0 523--532, 2013.
\newblock ISSN 0002-9939.
\newblock \doi{10.1090/S0002-9939-2012-11325-5}.
\newblock URL \url{https://doi.org/10.1090/S0002-9939-2012-11325-5}.

\bibitem[Berkes et~al.(2003{\natexlab{a}})Berkes, Horv\'{a}th, and Kokoszka]{Berkesetal2003a}
Istv\'{a}n Berkes, Lajos Horv\'{a}th, and Piotr Kokoszka.
\newblock Asymptotics for {GARCH} squared residual correlations.
\newblock \emph{Econometric Theory}, 19\penalty0 (4):\penalty0 515--540, 2003{\natexlab{a}}.
\newblock ISSN 0266-4666,1469-4360.
\newblock \doi{10.1017/S0266466603194017}.
\newblock URL \url{https://doi.org/10.1017/S0266466603194017}.

\bibitem[Berkes et~al.(2003{\natexlab{b}})Berkes, Horv\'ath, and Kokoszka]{Berkesetal2003b}
Istv\'an Berkes, Lajos Horv\'ath, and Piotr Kokoszka.
\newblock {GARCH processes: structure and estimation}.
\newblock \emph{Bernoulli}, 9\penalty0 (2):\penalty0 201 -- 227, 2003{\natexlab{b}}.
\newblock \doi{10.3150/bj/1068128975}.
\newblock URL \url{https://doi.org/10.3150/bj/1068128975}.

\bibitem[Billingsley(1995)]{Billingsley1995}
Patrick Billingsley.
\newblock \emph{Probability and measure}.
\newblock Wiley Series in Probability and Mathematical Statistics. John Wiley \& Sons, Inc., New York, third edition, 1995.
\newblock ISBN 0-471-00710-2.
\newblock A Wiley-Interscience Publication.

\bibitem[Box and Pierce(1970)]{BoxPierce1970}
G.~E.~P. Box and David~A. Pierce.
\newblock Distribution of residual autocorrelations in autoregressive-integrated moving average time series models.
\newblock \emph{Journal of the American Statistical Association}, 65\penalty0 (332):\penalty0 1509--1526, 1970.
\newblock ISSN 01621459.
\newblock URL \url{http://www.jstor.org/stable/2284333}.

\bibitem[Brockwell and Davis(1991)]{BrockwellDavis1991}
Peter~J. Brockwell and Richard~A. Davis.
\newblock \emph{Time series: theory and methods}.
\newblock Springer Series in Statistics. Springer-Verlag, New York, second edition, 1991.
\newblock ISBN 0-387-97429-6.
\newblock \doi{10.1007/978-1-4419-0320-4}.
\newblock URL \url{https://doi.org/10.1007/978-1-4419-0320-4}.

\bibitem[Davis et~al.(2018)Davis, Matsui, Mikosch, and Wan]{Davisetal2018}
Richard~A. Davis, Muneya Matsui, Thomas Mikosch, and Phyllis Wan.
\newblock Applications of distance correlation to time series.
\newblock \emph{Bernoulli}, 24\penalty0 (4A):\penalty0 3087--3116, 2018.
\newblock ISSN 1350-7265.
\newblock \doi{10.3150/17-BEJ955}.
\newblock URL \url{https://doi.org/10.3150/17-BEJ955}.

\bibitem[Durbin(1973)]{Durbin1973}
J.~Durbin.
\newblock \emph{Distribution theory for tests based on the sample distribution function}.
\newblock Society for Industrial and Applied Mathematics, 1973.
\newblock ISBN 0-89871-007-3, 978-0-89871-007-6.
\newblock \doi{10.1137/1.9781611970586}.
\newblock URL \url{https://epubs.siam.org/doi/abs/10.1137/1.9781611970586}.

\bibitem[Durbin(1976)]{Durbin1976}
J.~Durbin.
\newblock Kolmogorov-smirnov tests when parameters are estimated.
\newblock In Peter Gaenssler and P{\'a}l R{\'e}v{\'e}sz, editors, \emph{Empirical Distributions and Processes}, pages 33--44, Berlin, Heidelberg, 1976. Springer Berlin Heidelberg.
\newblock ISBN 978-3-540-37515-9.
\newblock \doi{10.1007/BFb0096877}.
\newblock URL \url{https://doi.org/10.1007/BFb0096877}.

\bibitem[Gaunt(2021)]{Gaunt2021}
Robert~E. Gaunt.
\newblock A simple proof of the characteristic function of {S}tudent's {$t$}-distribution.
\newblock \emph{Comm. Statist. Theory Methods}, 50\penalty0 (14):\penalty0 3380--3383, 2021.
\newblock ISSN 0361-0926.
\newblock \doi{10.1080/03610926.2019.1702695}.
\newblock URL \url{https://doi.org/10.1080/03610926.2019.1702695}.

\bibitem[Kulperger and Yu(2005)]{KulpergerYu2005}
Reg Kulperger and Hao Yu.
\newblock High moment partial sum processes of residuals in {GARCH} models and their applications.
\newblock \emph{Ann. Statist.}, 33\penalty0 (5):\penalty0 2395--2422, 2005.
\newblock ISSN 0090-5364,2168-8966.
\newblock \doi{10.1214/009053605000000534}.
\newblock URL \url{https://doi.org/10.1214/009053605000000534}.

\bibitem[Ljung and Box(1978)]{LjungBox1978}
G.~M. Ljung and G.~E.~P. Box.
\newblock {On a measure of lack of fit in time series models}.
\newblock \emph{Biometrika}, 65\penalty0 (2):\penalty0 297--303, 08 1978.
\newblock ISSN 0006-3444.
\newblock \doi{10.1093/biomet/65.2.297}.
\newblock URL \url{https://doi.org/10.1093/biomet/65.2.297}.

\bibitem[Pfister et~al.(2018)Pfister, Bühlmann, Schölkopf, and Peters]{Pfisteretal2018}
Niklas Pfister, Peter Bühlmann, Bernhard Schölkopf, and Jonas Peters.
\newblock Kernel-based tests for joint independence.
\newblock \emph{Journal of the Royal Statistical Society. Series B (Statistical Methodology)}, 80\penalty0 (1):\penalty0 5--31, 2018.
\newblock ISSN 13697412, 14679868.
\newblock URL \url{http://www.jstor.org/stable/44681792}.

\bibitem[Shaked and Shanthikumar(2007)]{ShakedShanthikumar2007}
Moshe Shaked and J.~George Shanthikumar.
\newblock \emph{Stochastic orders}.
\newblock Springer Series in Statistics. Springer, New York, 2007.
\newblock ISBN 978-0-387-32915-4; 0-387-32915-3.
\newblock \doi{10.1007/978-0-387-34675-5}.
\newblock URL \url{https://doi.org/10.1007/978-0-387-34675-5}.

\bibitem[Stephens(1978)]{Stephens1978}
Michael~A Stephens.
\newblock On the half-sample method for goodness-of-fit.
\newblock \emph{Journal of the Royal Statistical Society: Series B (Methodological)}, 40\penalty0 (1):\penalty0 64--70, 1978.
\newblock \doi{https://doi.org/10.1111/j.2517-6161.1978.tb01649.x}.
\newblock URL \url{https://rss.onlinelibrary.wiley.com/doi/abs/10.1111/j.2517-6161.1978.tb01649.x}.

\bibitem[Stephens(1986)]{Stephens1986}
Michael~A Stephens.
\newblock Tests based on edf statistics.
\newblock In Ralph~B. D'Agostino and Michael~A. Stephens, editors, \emph{Goodness-of-fit techniques}, volume~68 of \emph{Stat., Textb. Monogr.}, pages 97--194. Marcel Dekker, 1986.

\bibitem[Sz\'{e}kely et~al.(2007)Sz\'{e}kely, Rizzo, and Bakirov]{Szekelyetal2007}
G\'{a}bor~J. Sz\'{e}kely, Maria~L. Rizzo, and Nail~K. Bakirov.
\newblock Measuring and testing dependence by correlation of distances.
\newblock \emph{Ann. Statist.}, 35\penalty0 (6):\penalty0 2769--2794, 2007.
\newblock ISSN 0090-5364.
\newblock \doi{10.1214/009053607000000505}.
\newblock URL \url{https://doi.org/10.1214/009053607000000505}.

\bibitem[Wan and Davis(2022)]{DavisWan2020}
Phyllis Wan and Richard~A. Davis.
\newblock Goodness-of-fit testing for time series models via distance covariance.
\newblock \emph{Journal of Econometrics}, 227\penalty0 (1):\penalty0 4--24, 2022.
\newblock ISSN 0304-4076.
\newblock \doi{10.1016/j.jeconom.2020.05.008}.
\newblock URL \url{https://doi.org/10.1016/j.jeconom.2020.05.008}.

\end{thebibliography}

\end{document}